\documentclass[structabstract]{aa}  
\usepackage{graphicx}
\usepackage{txfonts}
\usepackage{aas_macros}
\usepackage{amssymb}
\usepackage{setspace}
\usepackage{latexsym}
\usepackage{color}
\usepackage{dcolumn}
\usepackage{bm}
\usepackage{subfig}
\usepackage{natbib, hyperref}  
\bibpunct{(}{)}{;}{a}{}{,} 

\begin{document}
   \title{Pushing the limits of the Gaia space mission\\ by analyzing galaxy morphology}
   \author{A. Krone-Martins\inst{1,2,3},
             C. Ducourant\inst{3},
             R. Teixeira\inst{2},
             L. Galluccio\inst{4},
             P. Gavras\inst{5},\\
             S. dos Anjos\inst{2}, 
             R. E. de Souza\inst{2},
             R. E. G. Machado\inst{2},
             \and
             J.-F. Le Campion\inst{3}
          }

   \offprints{algol@sim.ul.pt}

   \institute{$^{1}$SIM, Faculdade de Ci\^encias da Universidade de Lisboa, Ed. C8, Campo Grande, 1749-016 Lisboa, Portugal\\
			$^{2}$IAG, Universidade de S\~{a}o Paulo, Rua do Mat\~{a}o, 1226,  05508-900 S\~{a}o Paulo, SP, Brazil\\
			$^{3}$Observatoire Aquitain des Sciences de l'Univers, Laboratoire d'Astrophysique de Bordeaux, CNRS-UMR 5804, BP 89, 33271 Floirac Cedex, France\\
			$^{4}$Laboratoire Lagrange, UMR 7293, Universit\'e de Nice Sophia-Antipolis, CNRS, Observatoire de la C\^ote d'Azur, BP 4229, Nice Cedex 4, France\\
			$^{5}$GEPI, Observatoire de Paris, CNRS, Universit\'e Paris Diderot, 5 Place Jules Janssen, 92190 Meudon, France\\
}
   \date{Received ,  -----; accepted -----}

  \abstract
   {The ESA Gaia mission, to be launched during 2013, will observe billions of objects, among which many galaxies, during its 5 yr scanning of the sky. This will provide a large space-based dataset with unprecedented spatial resolution.}
   {Because of its natural Galactic and astrometric priority, Gaia's observational strategy was optimized for point sources. Nonetheless, it is expected that $\sim 10^6$ sources will be extragalactic, and a large portion of them will be angularly small galaxies. Although the mission was designed for point sources, an analysis of the raw data will allow the recovery of morphology of those objects at a $\sim0.2"$ level. This may constitute a unique all-sky survey of such galaxies. We describe the conceptual design of the method adopted for the morphological analysis of these objects as well as first results obtained from data simulations of low-resolution highly binned data.}
   {First, the raw Gaia 1D observations are used to reconstruct a 2D image of the object --  this image is known to contain artifacts and reconstruction signatures. Then, parameters characteristic of the reconstructed image are measured, and used for a purely morphological classification. Finally, based on the classification, a light profile is selected (pure-disk, disk+bulge, pure bulge, point source+disk, etc.) and fitted to all Gaia 1D observations simultaneously in a global process using forward modeling.}
   {Using simulations of Gaia observations from official Gaia Data Processing and Analysis Consortium tools, we were able to obtain the preliminary classification of the simulated objects at the $\sim83\%$ level for two classes (ellipticals, spirals/irregulars) or at the $\sim79\%$, $\sim56\%$, and $\sim74\%$ levels for three classes (E, S and I). The morphological parameters of simulated object light profiles are recovered with errors at the following levels: $-9\pm36\%$ for the bulge radius, $11\pm53\%$ for the bulge intensity, $1\pm4\%$ for the disk radius, and $-1\pm±7\%$ for the disk intensity. From these results, we conclude that it is possible to push the limits of the Gaia space mission by analysing galaxy morphology.}
   {}

 \titlerunning{Analysis of galaxy morphology with Gaia}
 \authorrunning{Krone-Martins et al.}
   \maketitle

\section{Introduction}

One of the most ambitious projects of modern astronomy, the Gaia mission, will observe more than one billion objects throughout the sky, providing astrometric, photometric, and spectroscopic data with unprecedented precision \citep{Perryman:2001p3838, Mignard:2008p9245, Lindegren:2010p9251}. Its launch is scheduled for 2013.

The satellite will work in scanning mode, systematically observing all objects brighter than magnitude G$\sim$20 (covering 330-1050 nm). The scanning law has been designed so that each object will be observed about 72 times on different scanning angles during the mission lifetime. Gaia was designed to explore the stellar content of the Galaxy with an unprecedented precision, and for this reason the mission has been optimized for point-like sources. Nevertheless, a large number of objects associated with an extended emission will be observed. From these, a majority will be angularly small galaxies with unresolved stellar content.

The simulation of the sky's content as observable by Gaia was recently presented in \cite{Robin:2012p9531}. In particular, it is predicted that about tens of millions of unresolved galaxies from the local Universe (z~$<0.3$) could be observed. The majority of these objects, however, will not be detected by Gaia. An onboard video processing algorithm (VPA) is in charge of performing an onboard selection of the data that will be transmitted to Earth. Criteria about the luminosity profile (stellar-like) of objects together with their apparent size and magnitude will be applied, thus rejecting a large part of the galaxies from transmission. In a parallel work, \cite{deSouza2013} assess the estimate number and characteristics of the galaxies observed by Gaia that should be transmitted to Earth. These authors conclude that about half a million of angularly small galaxies will be observed by Gaia and note that these will mostly be galaxies with a prominent bulge and/or nucleated galaxies, and that most spirals with faint bulges will be rejected onboard, as expected in \citet{Robin:2012p9531}. Thus, the major portion of detections will concern ellipticals and S0-Sb galaxies.

The galaxy content of our Universe is an active field of research, comprising several dedicated surveys. The best strategy for cosmological evolution studies is to perform high-resolution imaging, ideally space-borne, to resolve structural properties of galaxies as best as possible. This is the case of the Hubble Deep Field \citep{Williams1996}, which imaged a 5 arcmin$^{2}$ area, followed by GOODS \citep{Giavalisco2004}, which covered a larger area (360 arcmin$^2$), the Ultra Deep Field (UDF) survey \citep{Beckwith2006} or, more recently, the COSMOS survey \citep{scoville2007} designed to observe a 2 sq.deg. field mostly with the Hubble Space Telescope (HST) and Spitzer to probe galaxy evolution.

These surveys aimed at sampling volumes in the Universe in selected directions using multi-wavelength observations. The Sloan Digital Sky Survey \citep[][and refs. therein]{ahn2012} in contrast, is a ground-based survey covering about 1/4 of the sky and intended to study the large-scale structure seen in the distribution of galaxies and quasars. It provides spectroscopic redshift for about 1.5 million galaxies. However, its observations are Earth-based, and thus they are seeing-limited and do not allow a precise morphological analysis of angularly small objects.

The characteristics of Gaia are different from those of these deep extragalactic surveys, because it is an all-sky survey designed for the study of our Galaxy. However, owing to its high-resolution, it will observe the 1-2 arcsecs central region of angularly small galaxies with a pixel scale of 59 mas/pixel and a point spread function of $\sim$180 mas, up to magnitude G$\sim$20. These galaxies will be classified by the Data Processing and Analysis Consortium (DPAC) using photometry, spectroscopy, and astrometry \citep[e.g.][]{Tsalmantza2007, Tsalmantza2009}, and the high resolution of the satellite's observations combined with the appropriate treatment will allow one to derive the morphology of these objects.

This catalog will be biased toward the local (z~$\lesssim0.2$) angularly small elliptical galaxies and bulges, and will constitute the first whole-sky high-resolution study of such objects. The morphological analysis of these galaxies represents an insight into a population of the local Universe rarely imaged with such a high resolution (only a few thousand of them benefit from HST or adaptive optics observations) and never studied on such a large scale. This high-resolution local sample will deepen and clarify our vision of local bulges and spheroids, which was mostly built from photometric and structural studies. These objects are classically considered as stellar systems with a unique component and a smooth light distribution, containing an old stellar population and small amounts of gas and dust. Until the 1970s they were considered as relaxed, oblate systems, flattened by rotation and close to isotropic velocity dispersions \citep{Mo2011}. However, the past two to three decades of research have revealed a great complexity of their dynamical structure, star formation, and assembly histories \citep{Blanton2009}.

Today we know that elliptical galaxies can be roughly divided into three classes depending on their luminosity. The brightest ($M_B< -20.5$ mag) or giant ellipticals have higher S\'ersic indexes, boxy isophotes, less rotation support, more signs of triaxiallity, and core (shallow central surface brightness profile). On the other hand, ellipticals with intermediate ($-20.5 <M_B<-18$ mag) luminosity have disky isophotes, seem to be supported by rotation, and have steep central surface brightness cusps. Finally, the ellipticals with lower luminosity ($M_B> -18$ mag) or dwarfs reveal no or very little rotation and lower S\'ersic indexes. These differences probably indicate different physical mechanisms of formation and evolution.

This picture is roughly similar for the bulge component of spiral galaxies: classical bulges are believed to be formed through violent and fast processes, such as hierarchical clustering via minor mergers, which have and older stellar population; pseudo-bulges, through a longer time-scale secular evolution process, resulting in younger stellar populations; boxy-shaped bulges, which are characterized by cylindrical rotation. A refined analysis of larger and higher spatial resolution samples of these objects could greatly contribute to understanding their morphological evolution, and thus to studying the formation and evolution of structural components of the observed galaxies in the local and distant Universe.

The main objective of the present work is to describe how one can extract morphological information from Gaia observations of extended objects, that are expected to be dominated by these angularly small galaxies. For this purpose, we designed and implemented a conceptual data reduction system capable of dealing with the unique features of Gaia observations, such as the fact that they are one-dimensional. We are designing and implementing a completely automatic system to analyze Gaia's observations of galaxies with effective radii of a few arcseconds ($\lesssim2"$). This system is based on image reconstructions, image parameter measurements and classification, and finally a galaxy light profile fitting (also called galaxy bulge/disk decomposition) operating on the Radon space. In this paper we describe the conceptual design adopted for this system as well as results of its application to simulations of Gaia low resolution data.

The paper is organized as follows. In Sect. 2, we describe the Gaia data that will be used to morphologically analyze the galaxies. In Sect. 3, we present an overview of the pipeline under design in the Gaia Data Processing and Analysis Consortium (DPAC). In Sect. 4, we describe the image measurement method and results from simulations of Gaia observations. In Sect. 5, we describe the classification and results from a subset of galaxies from the Hubble Deep Field North and a set of simulations. In Sect. 6 we present the profile fitting method and its results on a set of simulations of Gaia's lowest resolution data. Finally, in Sect. 7 we present the conclusions of the current study.

\section{Gaia observations}
The analysis of extended sources is simplified by studying their two-dimensional signals, however, due to its specific and astrometric requirements, Gaia will not acquire such data.

The satellite will observe the sky in a continuous scanning, created by the coupling of the satellite's rotation and precession movements, that will last for about five years. This `scanning law' is optimized to guarantee that each point in the sky will be observed several times during the mission, while maintaining the satellite shielded from the Sun.

During the entire mission, the CCD charges in the focal plane are synchronously transferred to compensate the apparent sky motion and to allow integrating the observed objects. This requires a simultaneous reading of the CCDs. However, because Gaia's focal plane is very large, comprising 106 individual detectors\footnote{A diagram of Gaia's focal plane can be found in \citet{Jordi:2010p7353}.} and almost one gigapixel, it would be not practicable to transfer its entire content to the Earth. So, the signal from the focal plane is analyzed in real-time onboard to detect astronomical sources. When sources are detected, rectangular ``windows'' comprising only a few arcseconds around each source are created and transferred to the Earth.

Because Gaia's design was optimized for studying the stellar content of our Galaxy, which is mostly point sources, these observations will almost be one-dimensional. Moreover, the pixels of its CCDs are rectangular, measuring 59 by 177 mas -- their highest resolution is aligned with the direction in which the satellite scans the sky.\footnote{In Gaia's terminology, this direction is called along-scan (AL), while the perpendicular direction is called across-scan (AC).}

To analyze extended two-dimensional and not necessarily symmetric astronomical sources, images can be reconstructed from 1D observations by adopting some method as a starting point (the simplest one would be an observation stacking). We have shown that Gaia windows will provide enough information for the image reconstruction around sources anywhere in the sky \citep{AKM-004}, and some algorithms \citep[such as][]{Dollet2005, Harrison:2011p8791} were already proposed in the literature for performing this reconstruction.

\subsection{Windows used for extended-object analysis}

The satellite's focal plane comprises several columns of CCDs, which are specialized for certain measurements. For the image reconstruction and the morphological analysis of extended objects, there are two regions of the focal plane that are of special interest: the sky mappers (SM) and the astro fields (AF). The sky mapper consists in two columns of CCDs mainly aimed at the onboard detection of the sources. It transmits angularly large, but highly binned windows to the ground. Their information allows a global but low-resolution analysis of the object's extended emission.

The astro field is the core region of Gaia's focal plane for astrometry, comprising nine columns of CCDs aimed at multiplexing the measurements that will be used to derive the sources's astrometric parameters. The AF CCDs produce unbinned windows in the along-scan direction (AL), albeit with smaller angular sizes. Their information allows a more localized, but high-resolution analysis of the extended emission. 

Some characteristics of the windows transmitted to the Earth from the focal plane columns that will be used for the analysis of extended objects are shown in Table \ref{tab:window_sizes}. At each transit, one window of each type (SM, AF1, AF2, AF5, AF8) will be transmitted to the ground.

\begin{table}[!h]
	\begin{center}
	\begin{tabular}{|l|l|c|c|c|}
		\hline
		CCD  & Source's           &  Transmitted  & Angular \\
		column & G magnitude & window          & size\\
		                        &                 & (AL $\times$ AC) & \\
				  &                 &({\it samples}) & (") \\
		\hline
	  	SM   & $G<13.0$          & $40\times6$ & $4.72\times2.12$\\
	 	   & $G>13.0$               & $20\times3$ &$4.72\times2.12$\\
		\hline
		AF 1 & $G<13.0$          & $18\times6$ &$1.06\times2.12$\\
	 	   & $13.0<G<16.0$ & $12\times1$ &$0.71\times2.12$\\
		   & $G>16.0$          & $  6\times1$ &$0.35\times2.12$\\		
			   \hline
	   	AF& $G<13.0$          & $18\times12$ &$1.06\times2.12$\\
		 2, 5, 8    & $13.0<G<16.0$ & $18\times1$ &$1.06\times2.12$\\
		   & $G>16.0$          & $12\times1$ &$0.71\times2.12$\\
			\hline
	\end{tabular}
	\end{center}
	\caption{Windows transmitted to the Earth by the Gaia satellite used for image reconstruction. {\it Samples} is the Gaia terminology for binned pixels.}
	\label{tab:window_sizes}
\end{table}

Although these observations will be unbiased for point sources and data from all objects in the sky brighter than magnitude $G\sim20$\footnote{This is a broad passband, covering from 330nm to 1000 nm. The nominal transmission curve can be found at \citet{Jordi:2010p7353}.} will be transmitted to the ground, this will not happen for the extended objects. Gaia's onboard detection and windowing algorithm are based on the flux gradient, so galaxies with stepper gradients, such as pure bulges, have more chances of being windowed, and thus transferred to the ground, than galaxies with shallower gradients, such as pure disks. At this moment the biases introduced by the detection algorithm on the transferred objects are under analysis through numerical simulations.

These windows, which are almost one-dimensional because they are integrated in the across-scan (AC) direction (as can be seen from Table \ref{tab:window_sizes}), together with additional information about how they are geometrically placed on the sky (derived through the monitoring and reconstruction of the satellite's attitude), are the basis for the morphological analysis of galaxies with Gaia data.

 \begin{figure}
   \centering
   \includegraphics[width=0.8\linewidth]{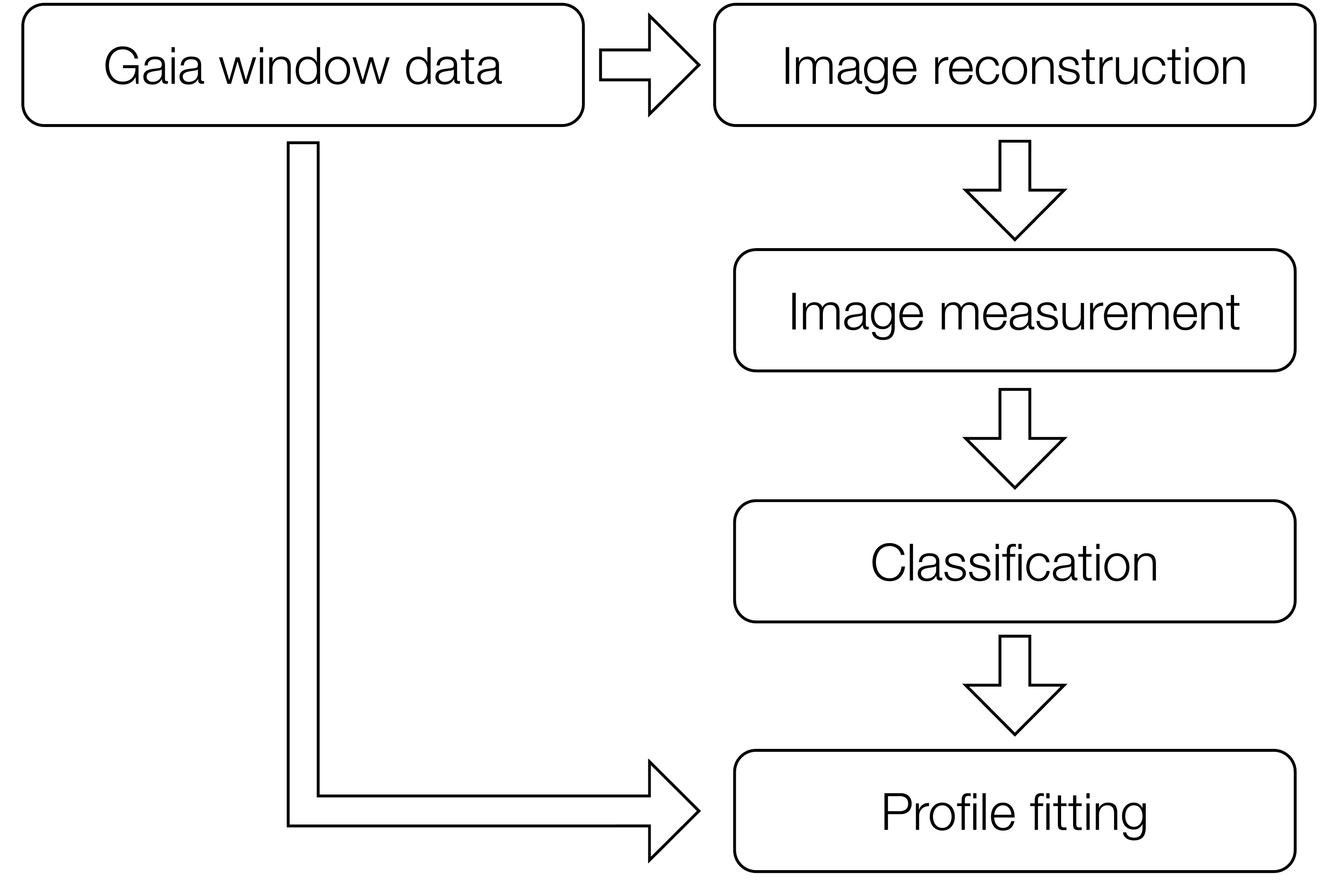}
      \caption{Diagram showing the major steps of the data reduction pipeline.}
         \label{fig:pipelinerepresentation}
   \end{figure}

\section{Pipeline overview}
To assess if it will be possible to extract morphological information from the transmitted windows, we have developed an automatic data reduction pipeline. Its main aim is analyzing the object's light profile through a bulge/disk decomposition, or a pure bulge or disk profile fitting depending on the morphological type of galaxy under analysis. This work is the framework for implementing the extended objects profile analysis pipeline of Gaia data under the Gaia Data Processing and Analysis Consortium.

The implemented pipeline works with a two-phases process, since it includes a supervised machine learning algorithm. This algorithm needs to be trained, so in a first phase the system is trained on a subset of the dataset, while in the second phase it is used to analyze the entire observational dataset. Both steps share an image reconstruction process as a starting point.

Since Gaia's ``almost 1D'' observations are very specific, no off-the-shelf astronomical analysis tool can be applied to its data processing. The design of our pipeline framework takes this into account. It also considers that the pipeline will deal with reconstructed images (which will certainly present some kind of reconstruction artifacts at different levels). Finally, it needs to be  efficient in terms of computing power, because $10^6$ galaxies are expected, and this analysis is not a mission priority. A representation of the pipeline steps can be found in Fig. \ref{fig:pipelinerepresentation}.

  \begin{figure*}
   \centering
  \begin{center}
  \subfloat[]{
    \includegraphics[width=0.3 \linewidth]{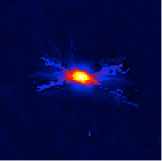}
  }
  \subfloat[]{
    \includegraphics[width=0.3 \linewidth]{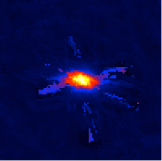}
  }
  \subfloat[]{
    \includegraphics[width=0.3 \linewidth]{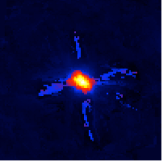}
  }
  \end{center}
      \caption{Example maps of ShuffleStack reconstructions of galaxies from simulations of Gaia one-dimensional data. An elliptical galaxy is presented in (a), a spiral in (b) and an irregular in (c). The 2D maps cover 3", with 30 mas/pixel. The objects were simulated with $G\sim18$, 0.798" major axes and an axis ratio equal to 2. Radial structures correspond to reconstruction artifacts.}
         \label{fig:exampleMapsShuffleStack}
   \end{figure*}

The first step is a two-dimensional image reconstruction based on the observed windows described in the last section. The most direct form of reconstruction would be the stacking and resampling of the different windows, taking into consideration the relative angles from which they were observed. However, in the past few years a certain number of studies have been conducted to allow a more precise image reconstruction, taking into consideration several characteristics of the Gaia observations, such as distinct binning and angular sizes among different windows and time-variable PSFs, as in \citet{Dollet2005}, \citet{Nurmi:2005p9057}, \citet{Pereyga2007}, and \citet{Harrison:2011p8791}.

The pipeline discussed in this paper is based on the currently available methods for image reconstruction, ShuffleStack (Harrison, priv. comm.)and FastStack \citep{Harrison:2011p8791}, even though both methods were created for point source reconstruction. Examples of reconstructions performed using the ShuffleStack method can be seen in Fig.~\ref{fig:exampleMapsShuffleStack}. The method to be used during the mission is still under discussion.

Once the reconstruction is performed, the pipeline classifies the object's morphology using a support vector machine. But, it is not the reconstructed image itself that is used to perform the classification, but some parameters measured on it. These parameters provide information about the flux distribution within the image and can be weighted by reconstruction error maps. This classification step is aimed to be purely morphological, as we are mainly interested to use its results to accelerate the fitting process, thus only light distribution dependent parameters were chosen.\footnote{In the future, we could also add other light distribution indicators, such as ellipticities.}

Finally, based on this preliminary classification, the object's light profile (pure-disk, disk+bulge, pure bulge) is selected and fitted simultaneously on all the available Gaia 1D windows in a global process using forward modeling. The artifacts observed in the 2D maps are created by the image reconstruction process and thus have no impact in the fitted parameters. If the fitting statistics of the profile are very poor, as may happen for a minor fraction of the objects, an alternative profile will be automatically fitted to the data.

In the following sections, the image measurement, classification, and profile fitting are described in more detail.

\section{Image measurement}
As the first step of the processing, a 2D map is reconstructed based on the satellite 1D window data. The main aim of this reconstruction is to perform object classification, which will be used to select the profile adopted in the forward model. However, the reconstructed pixels are not uniformly reliable. They depend on the reconstruction algorithm, but also on the  number and distribution of the overlapping windows at their position. They also depend on the resolution of each window. Therefore, in principle one should avoid feeding the pixel values themselves directly into the classification algorithms.

Our approach is to measure some image parameters that are able to provide clues about the object's morphological class without posing any strong constraint about the light profile of the object under analysis. The chosen parameters are the so-called CASGM20, which were used in the study of galaxy morphology when only a limited number of pixels was available for the analysis. They were introduced in \citet{Doi:1993p5085}, \citet{Abraham:1994p4126}, \citet{Abraham:1996p4217}, \citet{Abraham:2003p4894}, \citet{Conselice:2003p5139}, \citet{Conselice:2006p3903}, and \citet{Lotz:2004p5429}. We have shown in \cite{KroneMartins:2008p2436} that the computation of these parameters is quite robust under different binning conditions, and particularly under the sampling expected for reconstructed images from Gaia data. 

In this study, these parameters were measured on reconstructed 2D maps produced by methods that were not designed for extended sources. These methods performed point source reconstructions, and moreover, the resulting maps contain artifacts (as in Fig.~\ref{fig:exampleMapsShuffleStack}). Therefore, one should be cautious to physically interpret measurements of these parameters in reconstructed maps. Thus, P1-5 are used here to indicate them.

These parameters measure the light concentration in the center of the object (P1), the signal's asymmetry (P2), the smoothness or clumpiness of the light distribution in the image (P3), the concentration of light in pixels independently of their position in the image (P4), and the momentum of the brightest 20\% fraction of the flux within the image (P5). They can be expressed analytically as

\begin{equation} \label{eq:CASGM20}
\left(
 \begin{array}{ll} 
R_P(r) = \frac{\int_{0.8r}^{1.25r}dr'2\pi r'I(r')/[\pi(1.25^2-0.8^2)r^2]}{\int_0^rdr'2\pi r'I(r')/(\pi r^2)} & 
P1 = 5\log_{10}\left(\frac{r_{80}}{r_{20}}\right) \\
&\\
P2 = \frac{\sum_{i}\sum_{j} |I(i,j)-I_R(i,j)|}{\sum_{i}\sum_{j}I(i,j)} &
P3 = \frac{\sum_{i}\sum_{j} |I(i,j)-I_S(i,j)|}{\sum_{i}\sum_{j}I(i,j)} \\
&\\
P5 = \log_{10}\left(\frac{\sum_{l}M_l}{M_{tot}}\right), \mathrm{for}\  \sum_lI_l\le0.2I_{tot}  & 
P4 = \frac{\sum_{k}[2k-N-1]I_k}{ \langle I\rangle N(N-1)} \\
\end{array}
\right).
\end{equation}

In the equations above, $R_P(r)$ defines the Petrosian ratio\footnote{This is needed to compute the Petrosian radius $r_p$, defined to be the $r$ for which $R_p=0.2$. This radius is used in the computation of the C and S parameters.} as used in the SDSS \citep{Blanton:2001p6552}. $I(r)$ is the azimuthally averaged surface brightness profile. $r_{80}$ and $r_{20}$ are the radii that comprises $80\%$ and $20\%$ of the total flux of the galaxy (the flux inside $1.5r_p$). $M_l$ is the second-order momentum of the $l$-th pixel sorted by decreasing flux of the image. $M_{tot}$ is the second-order momentum computed for all pixels. $I_l$ if the flux of the $l$-th pixel. $I(i,j)$ is the flux of the image at the pixel $(i,j)$. $I_R(i,j)$ is the flux at $(i,j)$ of the image rotated by $180^\circ$. $I_k$ is the flux of the $k$-th pixel sorted by increasing flux. $N$ is the total number of pixels on the image. $I_l$ is the flux of the $l$-th pixel sorted by decreasing flux. And finally, $I_S(i,j)$ is the flux at $(i,j)$ of the image smoothed by a $0.3r_p$ boxcar filter.

 \begin{figure}
   \centering
   \includegraphics[width=1\linewidth]{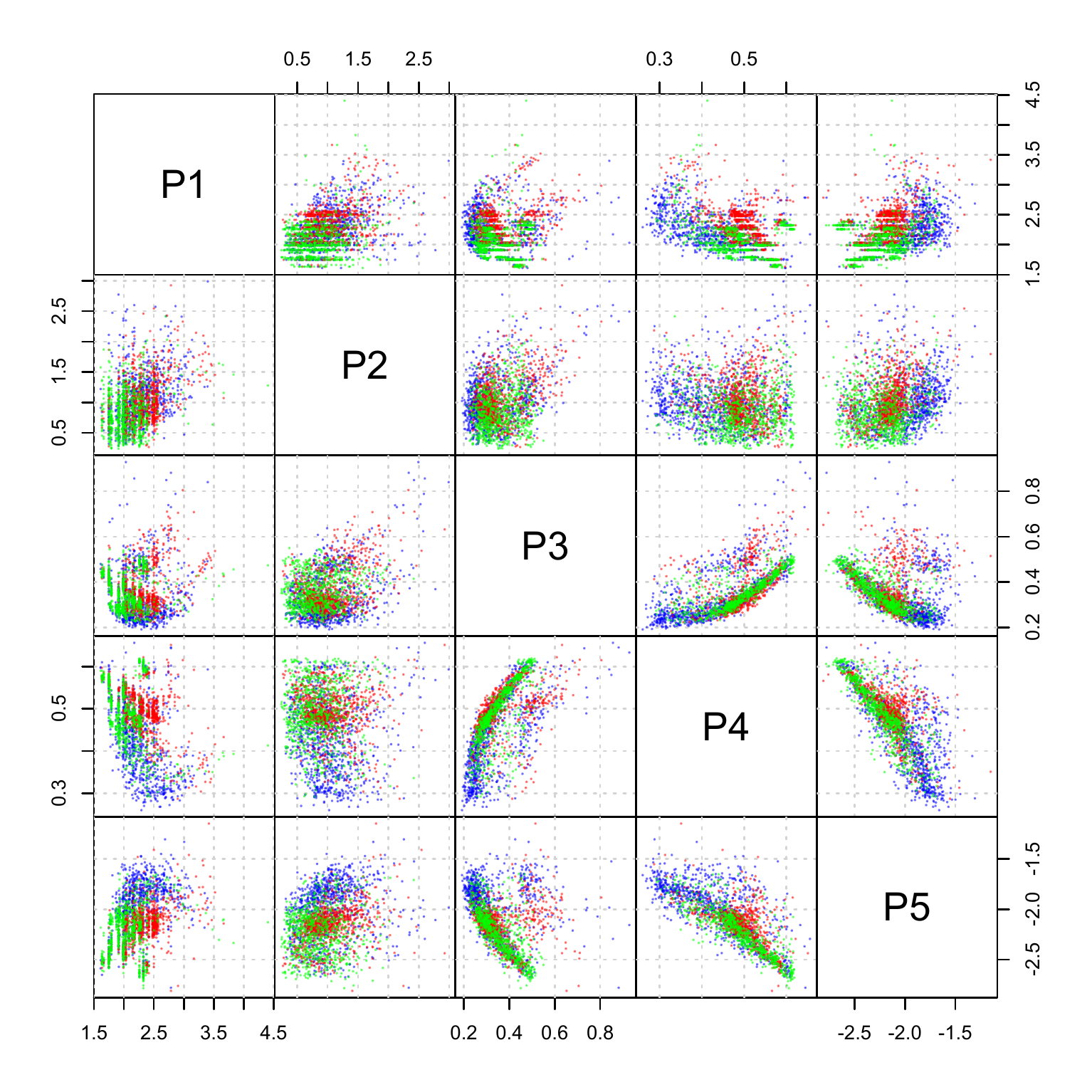}
      \caption{P1-5 parameters computed from reconstructed images using the ShuffleStack method for 2640 simulated galaxies of the elliptical (red), spiral (blue), and irregular (green) types.}
         \label{fig:shufflestack_casgm20space}
   \end{figure}

 \begin{figure}
   \centering
   \includegraphics[width=1\linewidth]{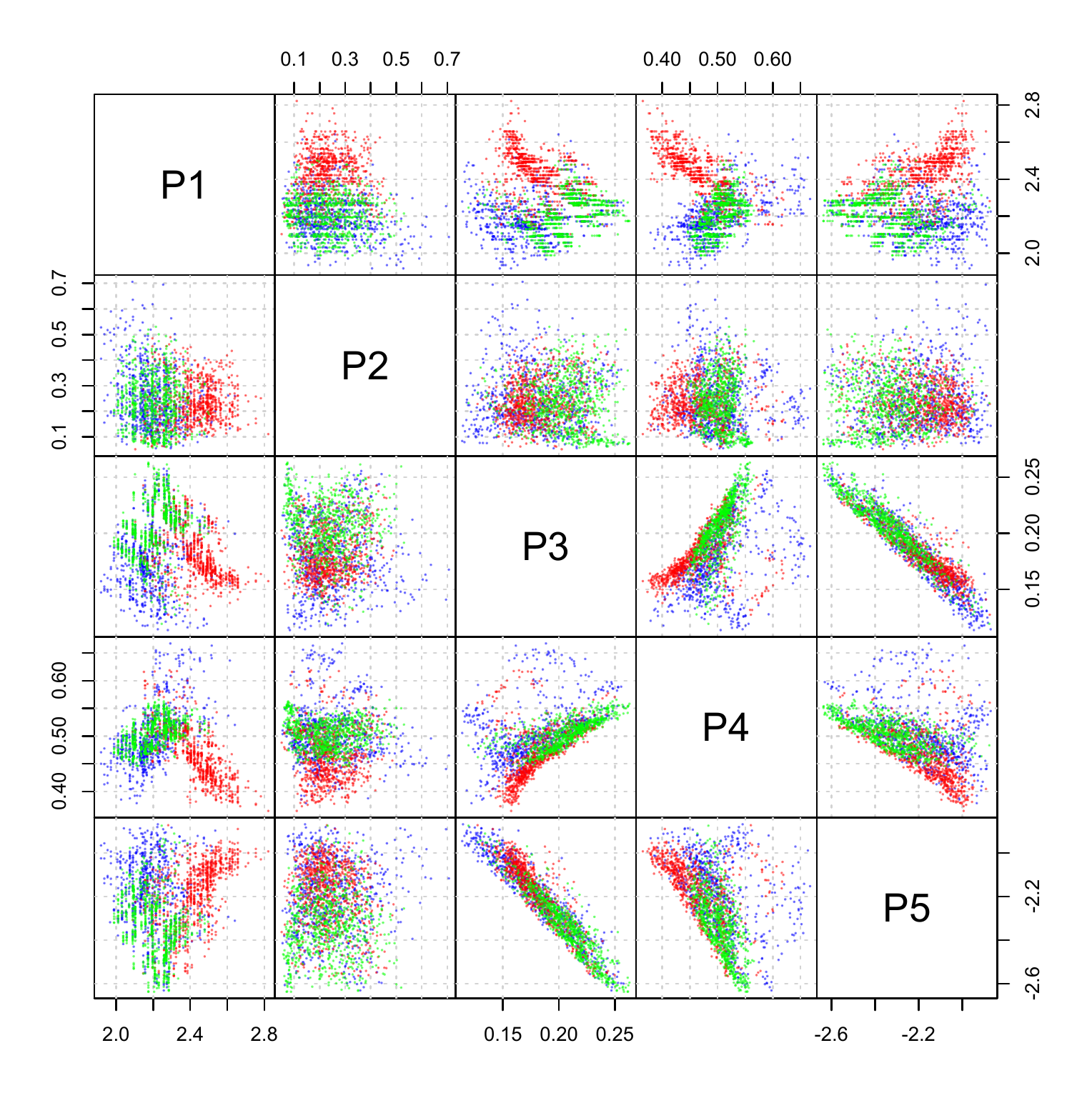}
         \caption{P1-5 parameters computed from reconstructed images using the FastStack method for 2640 simulated galaxies of the elliptical (red), spiral (blue), and irregular (green) types.}
         \label{fig:faststack_casgm20space}
   \end{figure}

One advantage of adopting the P1-5 parameters is that there is a straightforward way to take the error information from the image reconstruction process into account: a simple weighting scheme by the inverse of the pixel error can be used during their computation, thus allowing the uncertainty due to the reconstruction process to be taken into account during the classification step. However, because no reconstruction error map was available in the implementation of the image reconstruction methods, the tests were performed without weighting and thus represent a worst-case scenario. Moreover, we considered the background asymmetry to be null at this point in the processing, because the background of each observation used to reconstruct the signal and to measure the parameters is taken into account by a dedicated processing chain during the data reduction.

To check if the computation of these parameters on the reconstructed images had discriminatory capabilities for different galaxy classes, we simulated Gaia observations on a set of 2640 galaxies. This simulation was performed using Gaia's pixel-level simulator, the Gaia Instrument and Basic Image Simulator, or GIBIS, version 7 \citep{GIBIS7}, with MAGIL, which is the MAnager of Gaia Image Library \citep{Gavras_MAGIL_2010}.

The MAGIL code is responsible for introducing images of very well known objects into GIBIS, which are used as prototypes that are scaled in magnitude, size and axis-ratio to realistically simulate the desired object morphology on the focal plane. The Frei catalog \citep{Frei:1996p5214} is adopted as MAGIL's image library. Then, GIBIS performs all necessary simulations (PSF, scanning law, noise, TDI smearing, charge diffusion, radiation effects, charge-injections, windowing, binning, etc.) that result in the Gaia observations of the objects.

The simulations were performed in a region of the sky with a good observational coverage by Gaia windows, $(l,b) = (120^\circ; 45^\circ)$, where at most 82 transits will be available per object at the end of the mission. The simulated galaxies cover the G magnitude interval of [13; 20] mag, major-axis of [200; 1200] mas, axis-ratio [1; 10], and three morphological types (elliptical, spiral and irregulars), with 880 galaxies for each type.

The P1-5 parameters computed from the ShuffleStack reconstructed images are presented in Fig. \ref{fig:shufflestack_casgm20space}. It can be seen from that figure that although there is a strong overlap, different morphological types occupy distinct regions on all planes. These 2D integrated distributions of the five-dimensional P1-5 space indicates that a morphological classification based on these parameters is probably possible, if an adequate classifier is adopted.

To assess whether a different reconstruction algorithm would also enable morphological class discrimination, we employed the FastStack method \citep{Harrison:2011p8791}. The simulation of the window data was performed with GIBIS version 11 and MAGIL, and the characteristics of the simulated galaxies were the same as adopted for the ShuffleStack test. One can notice that the repartitions of the measured parameters differ from one reconstruction method to the other. This change is a result of the differences of the methods and their inherent artifacts. However, the results presented in Fig. \ref{fig:faststack_casgm20space} show that from these reconstructions, distinct galaxy types occupy different regions of the P1-5 space. This indicates that it is also possible to obtain a morphological discrimination among the different galaxy types purely based on window data with the FastStack method.

The distribution of the P1-5 parameters obtained with two independent reconstruction methods in Figs. \ref{fig:shufflestack_casgm20space} and \ref{fig:faststack_casgm20space} are very different, even though they are based on window data corresponding to the same simulated galaxies. This exemplifies that physical interpretations of these parameters measured from reconstructions performed with methods created for point sources should be avoided. In the present work, however, we are concerned with the discriminatory capability of these measurements, and thus with their application to object classification. The results obtained in this section indicate that a purely morphological classification is possible using Gaia observations, even if the method that will be used for the reconstruction during the final data treatment is still under discussion.

\section{Classification}
The classification step of our pipeline is intended to determine which profile will be fitted to the data. It relies on a supervised machine learning method called Support Vector Machine, or SVM \citep{Cortes:1995p9085, vapnik2000nature, CC01a}. This method is based on the separation of classes of objects using hyper-planes in a high-dimensional space, and the hyper-plane used for the classification is obtained through the maximization of its distance from the nearest training data points of the classes, which guarantees the lowest possible generalization errors. When the data under analysis are not linearly separable, as in the present work, a kernel function such as the radial basis function is adopted to map the data into a higher dimensional space, where they are linearly separable.

This algorithm was used before in astronomy, together with CASGM20 parameters, for galaxy classification in \citet{HuertasCompany:2008p5093}, and was tested on galaxy images under similar sampling conditions as those expected from Gaia reconstructed images in \citet{KroneMartins:2008p2436}. The CASGM20 parameters are known to contain redundant information, but SVMs are known to perform well in these conditions, and mixing SVM with strategies for dimensionality reduction and/or feature selection implies little or no improvement on the resulting classification accuracies \citep[see][]{Lin:2008p1505}, while it usually increases the computing time if the dimensionality is already low. Nonetheless, during the development of this work, we tested adopting of a Principal Component Analysis (PCA) before the SVM classification, and concluded that the results obtained for the classifications performed with and without the PCA step were compatible. To avoid penalizing the data processing with an additional step, we therefore decided not to adopt any dimensionality reduction technique.\footnote{We note, however, that when dealing with many more dimensions, a PCA should be considered.}

The SVM is a supervised algorithm, meaning that its results depend on the choice for the training strategy. Regarding our specific case, there are three possible ways to create the training set for the SVM algorithm: 

\begin{itemize}
\item adopt the classification obtained from the Gaia spectrophotometric data and semi-empirical models for a subset of the observed objects;
\item adopt a pre-classified input list of objects that will be observed by Gaia but that already have known morphological classes;
\item perform the fitting of all profile classes on high S/N sources, select the best-fitted profile as the object's class and train the SVM with the resulting object-class pairs.
\end{itemize}

In all these options, the training data are balanced between the different classes to reduce possible biases during the construction of the training set. In the first case, this is done by simply selecting the same amount for each type of object from Gaia's main database. In the second case, the input list is externally built by selecting about the same amount of objects in each class. In the third case, after the high S/N sources are fitted, the final training list is constructed by selecting an equilibrated amount of objects of each type.

Each of the training options described above have different advantages and drawbacks. The first strategy, although easily implementable, would prevent the interesting {\it a posteriori} comparison of the classifications obtained from color data with morphology data for the same objects with the same space mission. The second strategy faces the problem that there is no large number of objects already observed at the same spatial resolution as is expected from Gaia data, which will also be observed by Gaia. However, this option would create an external consistency with already known objects, acting as an external calibration for the unknown objects, and would probably also allow creating of a list with objects widely spread on the S/N dimension.

The advantage of the third, and last, option, would be that it produces a highly internally consistent classified set. However, because only high S/N objects would be used to perform the training, this could bias the classifier, since their light distribution characteristics are not necessarily the same as those of the faintest objects. Finally, this last option would be inefficient from the point-of-view of the required computing power, since the fitting may be the most time-consuming part of the processing, and it would run at least as many times as there are different light profiles for each object in the training set list.

The SVM training strategy that will be adopted for the final data reduction will be decided during the mission. This decision was taken since the nature of the satellite's real data can be different from the numerical simulations, and because the classification is an important step in the pipeline, it will be carefully analyzed on the face of real data.

In the next two subsections we apply this part of our pipeline to different datasets to validate and test it under different conditions. We note that during the mission data reduction, the training of the classifiers will be performed entirely based on Gaia data itself, and no simulations or external catalog data will be adopted.

\begin{figure}[!htb]
  \begin{center}
  \subfloat{
  \includegraphics[height=0.4\linewidth, trim=0cm 0.5cm 0.5cm 2cm, clip=true]{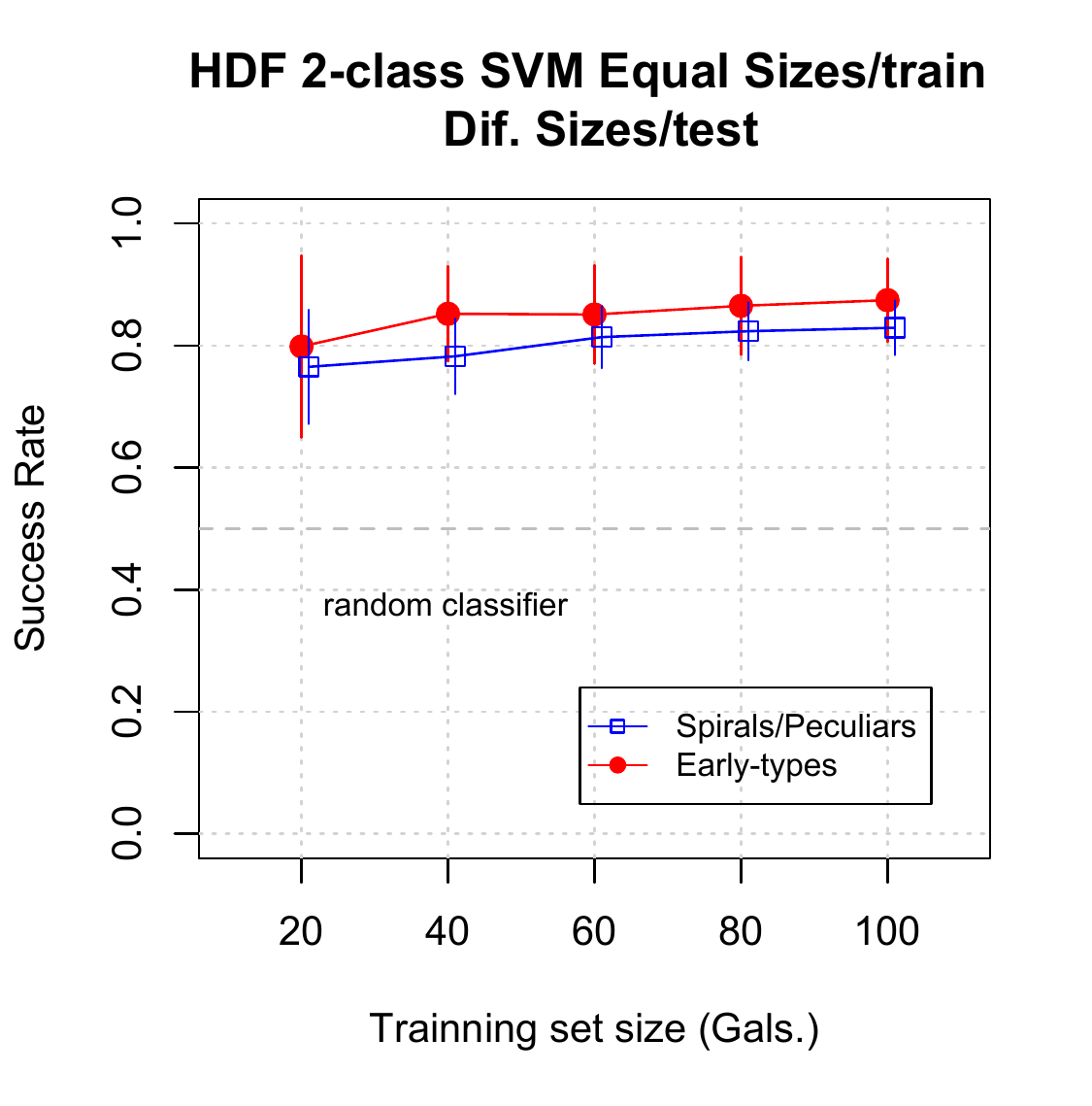}
  }
  \subfloat{
  \includegraphics[height=0.4\linewidth, trim=0cm 0.5cm 0.5cm 2cm, clip=true]{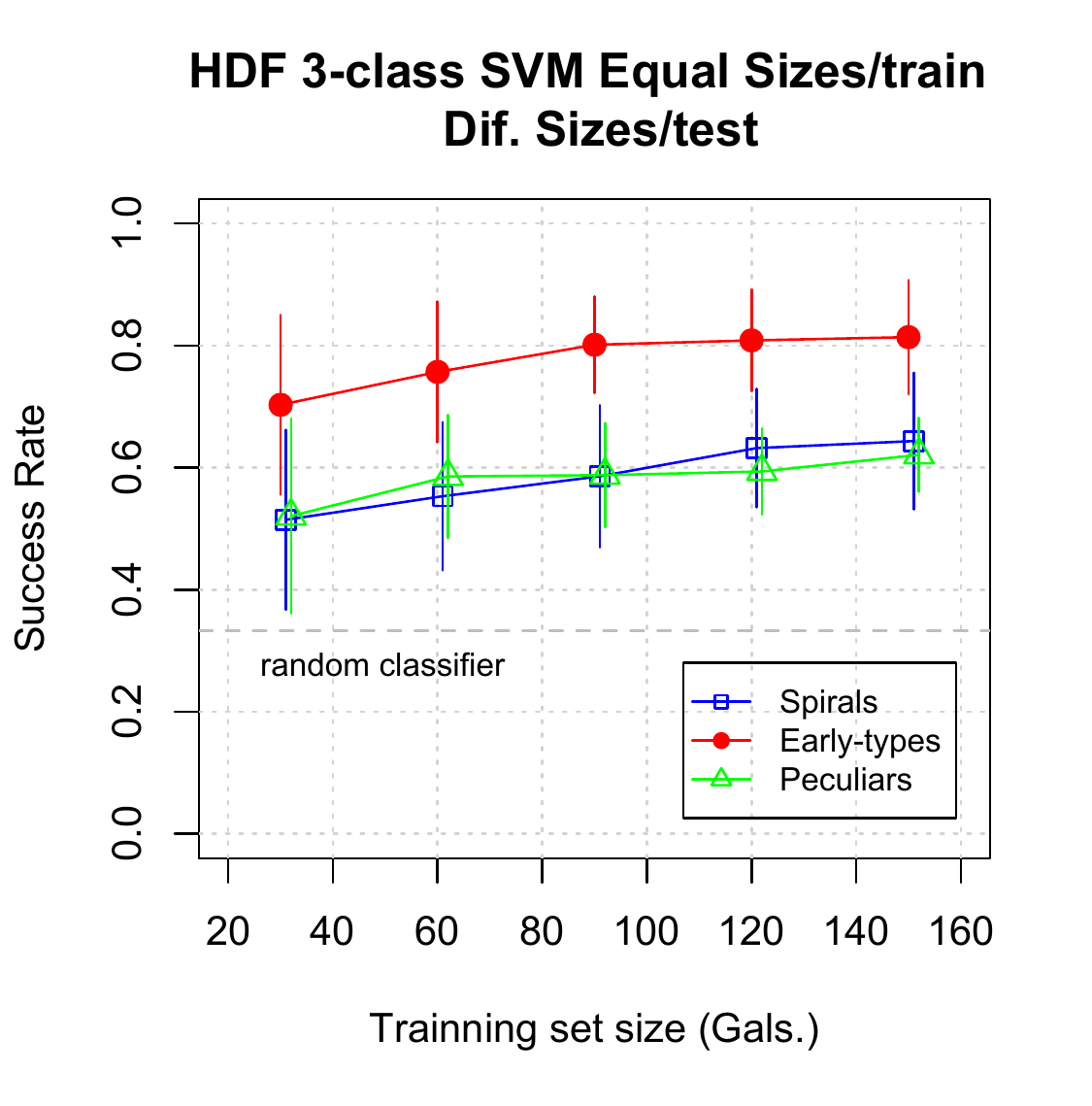}
  }
  \end{center}
  \caption{Classification success rate as a function of the training set size obtained with the classifier applied on HST-N data. The classification was performed in two and three classes.}\label{fig:casgm20_class_HDF}
\end{figure}

\begin{table}[!h]
	\begin{center}
	\begin{tabular}{|l|c|c|c|}
		\hline
Real class&Classified as Early&Class. as S or I\\
		\hline
		\hline
Early&$0.87\pm0.07$&$0.13\pm0.07$\\ 
S or I&$0.17\pm0.05$&$0.83\pm0.05$\\
		\hline
	\end{tabular}
	\end{center}
	
		\begin{center}
	\begin{tabular}{|l|c|c|c|}
		\hline
Real class&Class. as Early&Class. as S&Class. as I\\
		\hline
		\hline
Early&$0.81\pm0.09$&$0.27\pm0.07$&$0.07\pm0.06$\\  
S&$0.17\pm0.08$&$0.64\pm0.11$&$0.19\pm0.11$\\
I&$0.09\pm0.04$&$0.29\pm0.06$&$0.62\pm0.06$\\
		\hline
	\end{tabular}
	\end{center}
	\caption{Confusion matrix for the classification of the HDF-N data in two or three morphological classes (50 galaxies/type were used for the training).}
	\label{tab:confusion_hdf_2class}
\end{table}

\subsection{Validation on Hubble data}
To validate our classifier on real data, we tested it on a subset of the Hubble Deep Field North galaxies \citep{Williams1996}. The galaxies observed on this field will not be observed by Gaia because of the satellite's limiting magnitude. However, in its focal plane, Gaia observations of galaxies will present approximately the same sampling conditions and PSF scale as HDF-N\footnote{However, Gaia's PSF form is significantly different because of its rectangular primary mirrors.}. In this way, HDF-N could be considered as a test case for the classifier when it is applied to small galaxies.

We started by creating a four-color integrated image from the co-addition of HDF-N observations with the F300, F450, F606, and F814 filters. Then, galaxy stamps were extracted from this integrated image. The 400 brightest galaxies were selected to allow a more accurate visual classification, and were independently visually classified by two of the authors. From the resulting classifications, the 281 galaxies for which the result of both classifications was consistent, were selected for testing the automatic classifier.

Using a Monte Carlo approach, we randomly selected galaxies from this list to create pairs of training/testing lists. These pairs were produced 100 times, and the lists were used to train and test our classifier through determining the classifier's accuracy. The complete procedure was repeated for several training set sizes, and the results for the success rate\footnote{The success rate is equivalent to the true positive rate.} are presented in Fig. \ref{fig:casgm20_class_HDF}, while the confusion matrix is presented in Table \ref{tab:confusion_hdf_2class}.

These results show that the classifier distinguishes the elliptical galaxies more efficiently: it is possible to distinguish two classes, between early-types (ellipticals+lenticulars) and spiral+peculiar (including irregulars, interacting, etc.) types with an $\sim 85\%$ accuracy, and three classes, between early-types, spirals and peculiars with $\sim 81\%$, $\sim 64\%$, and $\sim 62\%$  accuracy.

They also indicate an increase on the success rate when a larger number of galaxies is available for the training, as expected from a supervised learning algorithm. This behavior is, nonetheless, asymptotic and small increments in the training set size would not significantly improve the results, while it would increase the training time. Adopting a simple logarithmic extrapolation law for the accuracy growth curve of Fig.~\ref{fig:casgm20_class_HDF} an almost six-folded increase in the training set size would be needed to reach $\sim 90\%$ accuracy level in the two-class classification, improving the results in less than $7\%$.

To test the classifier on a larger dataset, we performed tests using a sample of 3000 galaxy images from the ACS catalog of the COSMOS survey \citep{Leauthaud2007, scoville2007}. The data had previous automatic classification based on the HST-ACS images from Tasca Morphology Catalog\footnote{{http://irsa.ipac.caltech.edu/data/COSMOS/tables/morphology/}.} v1.1 \citep{Tasca2009}. This was the classification adopted as the underlying reference for our test. From the objects classed similarly by two independent methods (``dist\_int'' and ``class\_linee''), and with half-light radii $<4"$, the brightest 1000 galaxies of each class (early-types, spirals, irregulars) were selected. Objects containing saturated pixels were not considered, and 600 objects/class were used for training, while the rest was used for testing the classifier. The results indicated that using the COSMOS data our classifier reaches a success rate of $\sim 91\%$ for two classes, and $\sim 86\%$, $\sim 82\%$ and $\sim 90\%$ for three classes. The better separation between spirals and irregulars obtained using COSMOS data compared with HDF-N data might be due to a cleaner reference classification: while the underlying reference for COSMOS was created by the intersection of two automatic methods, the HDF-N test adopted visual classifications, which are inherently more uncertain. On the other hand, albeit adopting a more than ten-times larger training set than the test performed with HDF-N data, the improvement in early/late type classification was $\lesssim6\%$.

The results obtained from Hubble data discussed in this subsection could be considered as representative of a test case for the classifier when the images analyzed are free from artifacts, but when the training set is not perfectly built. They demonstrate that for a similar amount of information as expected from Gaia reconstructed images it is possible to obtain a good accuracy on the morphological classification between two types and a fair accuracy between three types.

\subsection{Results from simulated data}
To test our classifier under mission-like conditions, Gaia simulations performed with GIBIS and MAGIL and reconstructions were adopted. The reconstructions were produced by the ShuffleStack method applied to GIBIS 7 simulations and FastStack applied to GIBIS 11 simulations. They included 2640 simulated galaxies from three morphological types (elliptical, spiral, irregular). After the reconstructions, the P1-5 parameters were measured and used to train and test the classifier. These are the sets of parameters presented in Figs.~\ref{fig:shufflestack_casgm20space} and~\ref{fig:faststack_casgm20space}.

\begin{figure}[!htb]
  \begin{center}
  \subfloat{
  \includegraphics[height=0.4\linewidth, trim=0cm 0.5cm 0.5cm 2cm, clip=true]{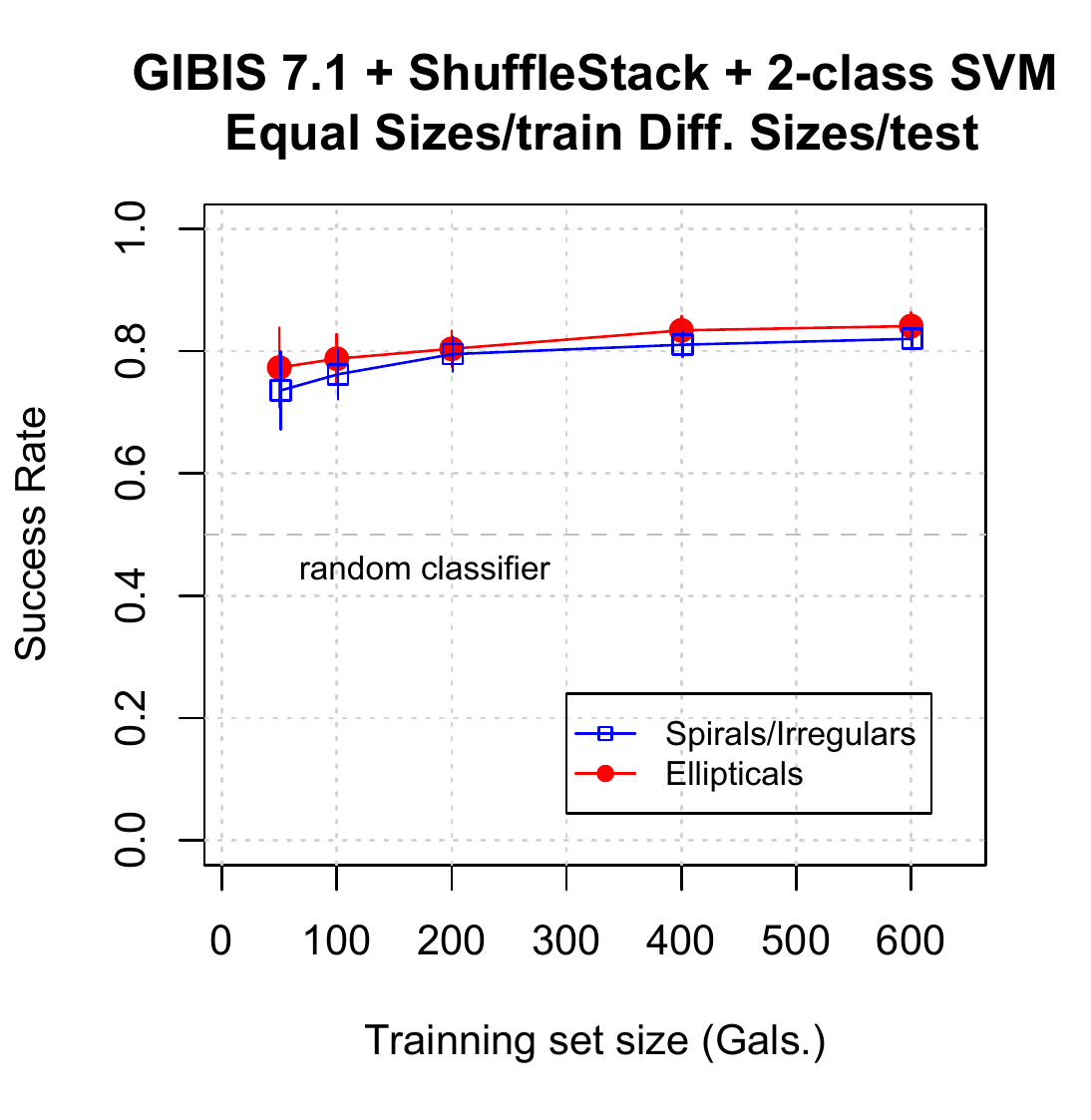}
  }
  \subfloat{
  \includegraphics[height=0.4\linewidth, trim=0cm 0.5cm 0.5cm 2cm, clip=true]{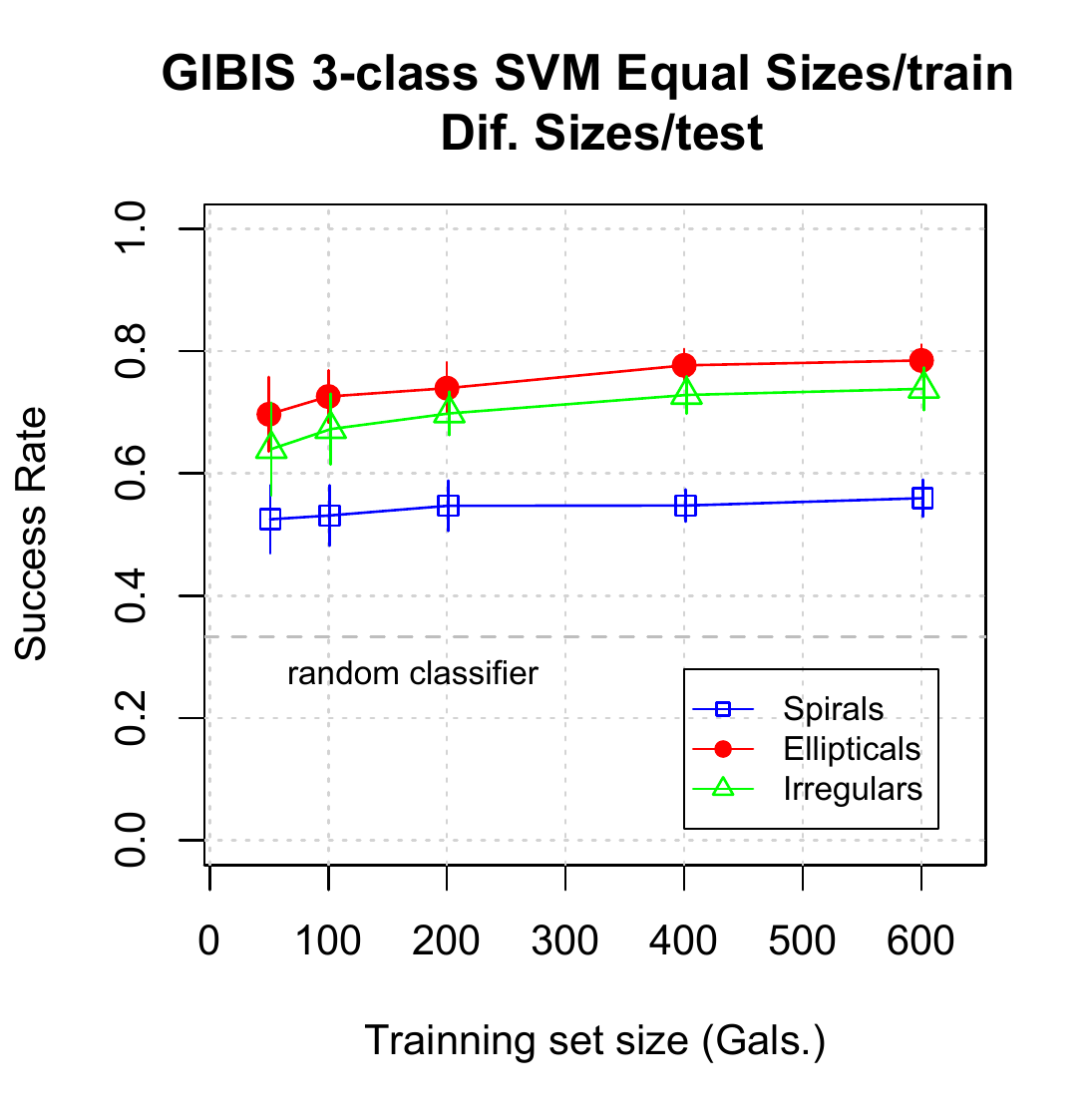}
  }
  \end{center}
  \caption{As Fig. \ref{fig:casgm20_class_HDF}, but for simulations reconstructed with ShuffleStack.}\label{fig:casgm20_class_GIBIS7_3types}
\end{figure}

\begin{figure}[!htb]
  \begin{center}
  \subfloat{
  \includegraphics[height=0.4\linewidth, trim=0cm 0.5cm 0.5cm 2cm, clip=true]{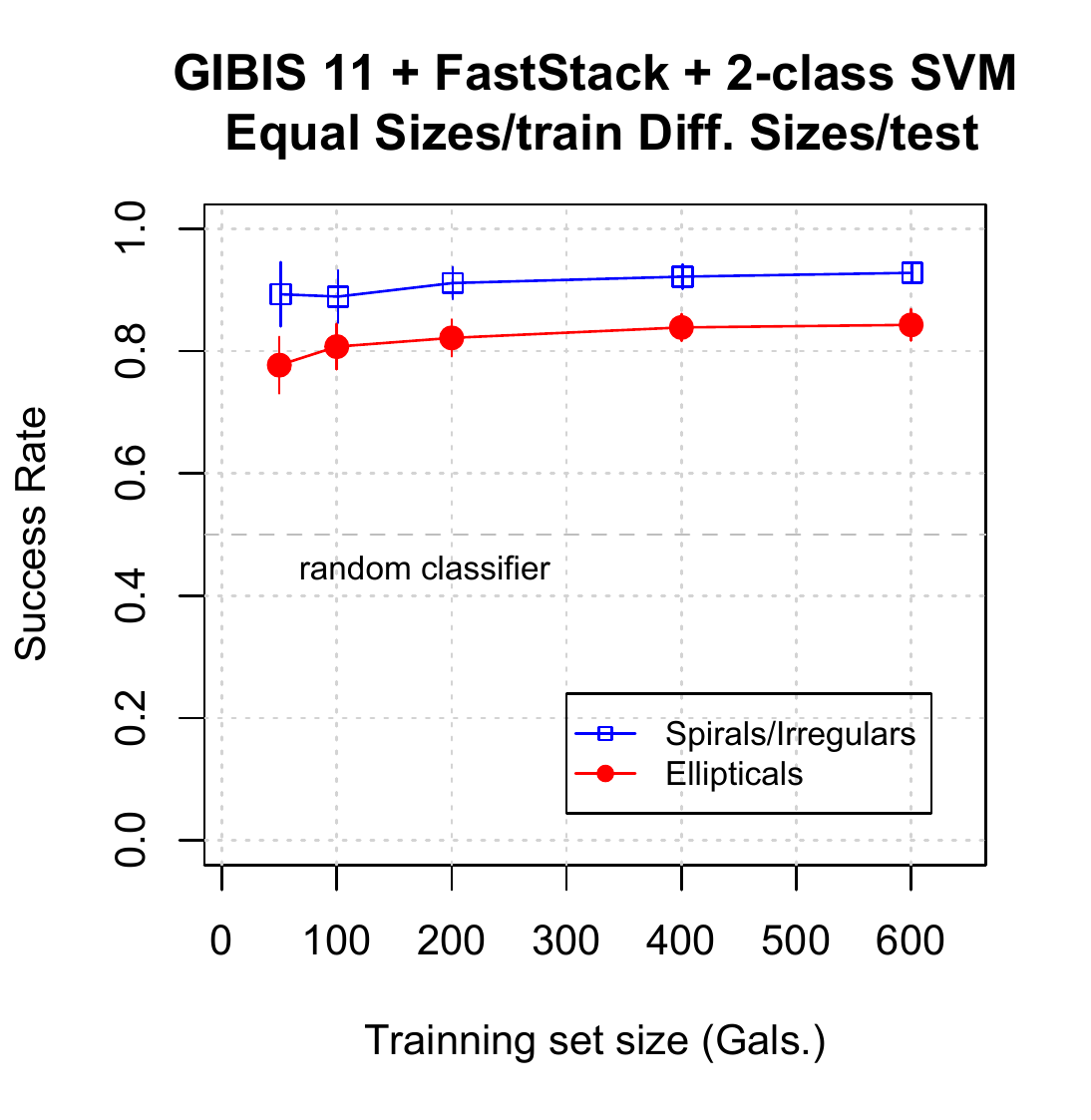}
  }
  \subfloat{
  \includegraphics[height=0.4\linewidth, trim=0cm 0.5cm 0.5cm 2cm, clip=true]{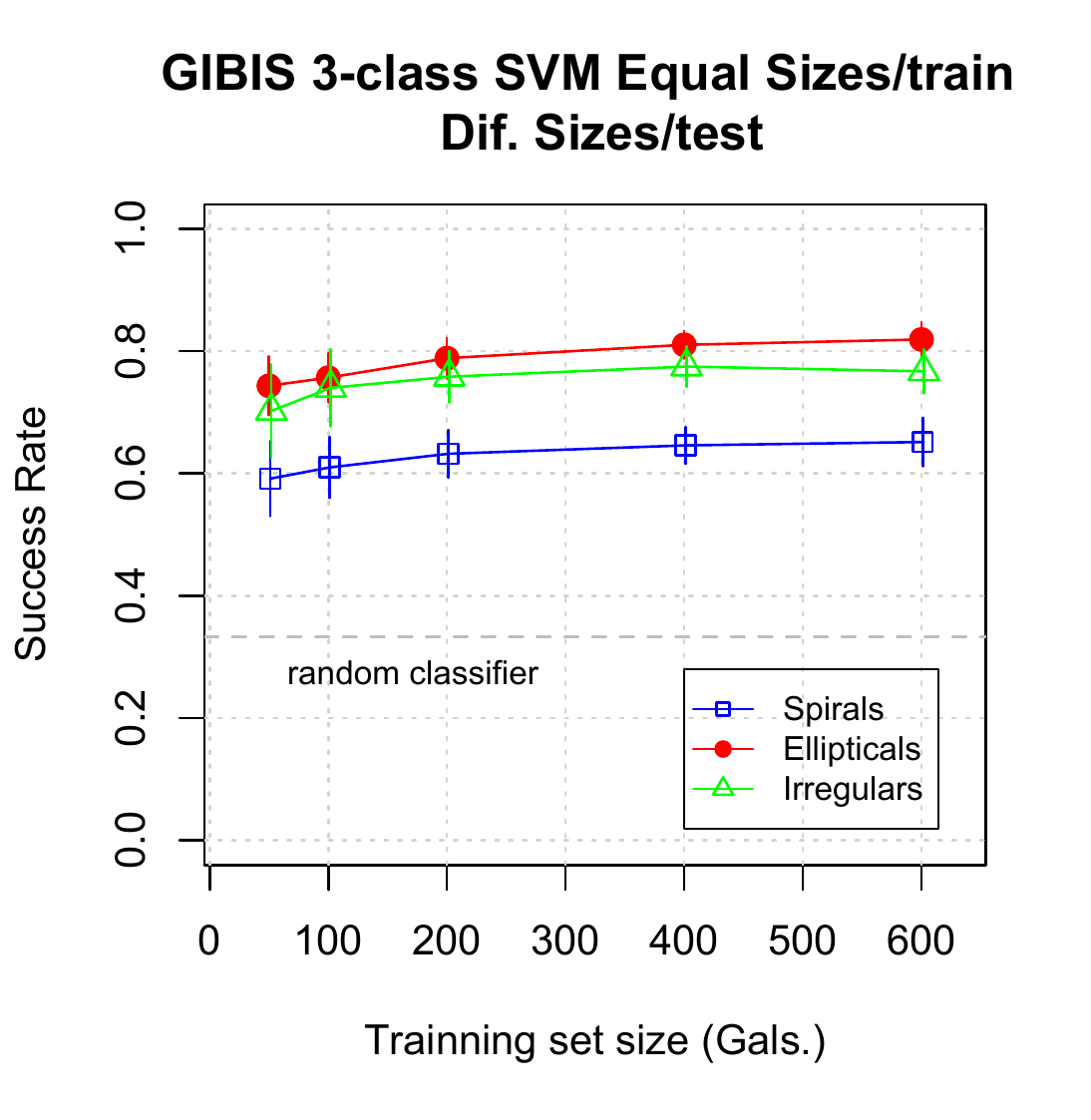}
  }
  \end{center}
  \caption{ As Fig. \ref{fig:casgm20_class_HDF}, but for simulations reconstructed with FastStack.}\label{fig:casgm20_class_GIBIS11_3types}
\end{figure}

\begin{table}[!h]
	\begin{center}
	\begin{tabular}{|l|c|c|c|}
		\hline
Real class&Classified as E&S or I\\
		\hline
		\hline
E&$0.84\pm0.02$&$0.16\pm0.02$\\
S or I&$0.18\pm0.02$&$0.82\pm0.02$\\
		\hline
	\end{tabular}
	\end{center}
	
		\begin{center}
	\begin{tabular}{|l|c|c|c|}
		\hline
Real class&Class. as E&Class. as S&Class. as I\\
		\hline
		\hline
E&$0.79\pm0.03$&$0.08\pm0.02$&$0.14\pm0.02$\\
S&$0.14\pm0.02$&$0.56\pm0.03$&$0.30\pm0.03$\\
I&$0.11\pm0.02$&$0.15\pm0.03$&$0.74\pm0.04$\\
		\hline
	\end{tabular}
	\end{center}
	\caption{Confusion matrix for the classification of the GIBIS data reconstructed with ShuffleStack in two or three morphological classes (600 galaxies per type were used for the training).}
	\label{tab:confusion_gibis7_shuffle_3class}
\end{table}
Using a similar procedure as in the last subsection, we constructed random resamplings of training/testing lists pairs using different sizes for the training set. These sets were used to test the success rate as well as confidence intervals for the classifier. The results for ShuffleStack and FastStack are presented in Figs.~\ref{fig:casgm20_class_GIBIS7_3types} and~\ref{fig:casgm20_class_GIBIS11_3types}. These results show that the success rate increases as the training set size increases, although this happens slowly.

As in the last subsection, we also built the confusion matrices for the largest training set, in this case comprising 600 galaxies per type for the two or three classifications classes. These matrices are presented in Tables \ref{tab:confusion_gibis7_shuffle_3class} and \ref{tab:confusion_gibis11_faststack_3class}.

The results for the two classes classification in Fig. \ref{fig:casgm20_class_GIBIS7_3types} and Fig. \ref{fig:casgm20_class_GIBIS11_3types} show that a success rate for two classes can be obtained at $>80\%$ levels using 200 galaxies per type. This high differentiation capability shows that bulge+disk or pure-bulge profiles can be correctly selected for the respective galaxies under analysis early in the mission. Although the galaxy detection performance of Gaia's video-processing algorithm is still being assessed, bulge+disk or pure-bulge profiles are the two types that we predict that will be most often detected and transferred to the Earth. Moreover, the results obtained based on ShuffleStack and FastStack reconstructions are compatible over all confusion matrices, as can be seen in Tables \ref{tab:confusion_gibis7_shuffle_3class} and \ref{tab:confusion_gibis11_faststack_3class}, albeit FastStack presents a systematic improvement. These classification results indicate that whatever reconstruction method is adopted, the training set needs to be constructed from 600 galaxies per type or more to reach a fair accuracy in the classification in three classes.

\begin{table}[!h]
	\begin{center}
	\begin{tabular}{|l|c|c|c|}
		\hline
Real class&Classified as E&S or I\\
		\hline
		\hline
E&$0.84\pm0.03$&$0.16\pm0.03$\\
S or I&$0.07\pm0.02$&$0.93\pm0.02$\\
		\hline
	\end{tabular}
	\end{center}
	
		\begin{center}
	\begin{tabular}{|l|c|c|c|}
		\hline
Real class&Class. as E&Class. as S&Class. as I\\
		\hline
		\hline
E&$0.82\pm0.03$&$0.07\pm0.02$&$0.11\pm0.03$\\
S&$0.06\pm0.02$&$0.65\pm0.04$&$0.29\pm0.04$\\
I&$0.04\pm0.02$&$0.19\pm0.03$&$0.76\pm0.04$\\
		\hline
	\end{tabular}
	\end{center}
	\caption{As Table~\ref{tab:confusion_gibis7_shuffle_3class}, but for simulations reconstructed with FastStack.}
	\label{tab:confusion_gibis11_faststack_3class}
\end{table}

From the results obtained with classifications performed from  ShuffleStack and FastStack reconstructions and from those obtained from real Hubble images, it is possible to argue that there is still some room to improve the reconstruction methods. However, the reconstruction method is not the limiting factor at the classification for two-class or three-class discrimination, because accuracies obtained from simulations and reconstructions, and from real Hubble images, are compatible.

The main conclusion obtained from this test is that using the proposed methods, a fair morphological classification can be obtained from realistic simulations of galaxy observations by Gaia that went through reconstruction methods that were not designed for extended sources, and thus, that a fair purely morphological classification can be expected based on Gaia observations.

\section{Profile fitting}
The final and central part of the analysis is fitting the brightness profiles on the Gaia 1D window data. The broad form of the light profiles of galaxies is known and can be decomposed into two main contributions: the disk and the bulge\footnote{Our implementation is based on a fully numerical approach, so additional contributions from central point sources, bars, spirals, etc. can easily be introduced, if needed.}. The most valuable information that can be obtained from Gaia data about the morphology of the galaxies that will be observed, are the parameters (radius, intensity, etc.) of these two components.

The light profile of these components are parametrized with an exponential profile for the disk and a S\'ersic one for the bulge. Using this parametrization, elliptical galaxies would be fitted with a S\'ersic profile, spirals with a composite profile, created from the addition of the exponential and S\'ersic profiles, and late spirals and irregulars with an exponential profile. For the irregulars, the obtained solutions would likely present poor fitting statistics. The profiles adopted for the bulge (S\'ersic) and disk (exponential) have the form

\begin{equation}
I_d(r)=I_{0d}\exp{\left(-\frac{r}{r_d}\right)},\\
I_b(r)=I_{0b}\exp{\left(-b_n\left[\left(\frac{r}{r_b}\right)^{\frac{1}{n}} - 1\right]\right),}
\end{equation}

\noindent where $I_d(r)$ is the light intensity of the disk at the radius $r$, $I_{0d}$ is the light intensity in the center of the disk, $r_d$ is the disk scale length, $r_b$ is the bulge effective radius, $I_{0b}$ is the bulge intensity at the effective radius, $n$ is the S\'ersic index, $b_n$ is a constant, and $r$ is the distance where the intensity is computed.

For the sake of the computational efficiency, we computed $b_n$ using the analytical approximation from \cite{Capaccioli:1989p7121} \cite[apud][]{Graham:2005p7165}, $b=1.9992n-0.3271$, which is valid for a broad range of S\'ersic indexes ($0.5 < n < 10$).

In addition to the usual parameters, such as radii, intensities, elipticities and position angles, the galaxies are also allowed to have diskness/boxiness distortions in our model, in the sense of \cite{Athanassoula:1990p6909}. This is implemented with the following transformation:

\begin{equation}
r(x,y)=\left[|x-x_{gal}|^{C_0+2}+\left|\frac{y-y_{gal}}{1-e}\right|^{C_0+2}\right]^{\frac{1}{C_0+2}},
\end{equation}

\noindent where $e$ is the ellipticity and $C_0$ represents the diskness/boxiness: $C_0=0$ is a normal ellipse, $C_0<0$ forms a diamond-like galaxy, $C_0>0$ forms a box-like galaxy.

The fitting of the Gaia window data is performed using forward modeling, where simulations of Gaia observations will be compared with all available real data simultaneously. To produce simplified simulated observations, we rely on the Radon transform\footnote{Although this transform is not widely known in astronomy, it was used before in \cite{Aime1978}, \citep{Zhang1993}, \cite{Touma1997}, \cite{Starck:2003p3898}, and more recently in \cite{Case2009}.}.

In the $\mathbb{R}^2$, the Radon transform \citep{Radon1917}\footnote{English translations in \cite{Deans1983} and \cite{Parks1986}.}, of a certain function $f(x,y)$ is defined as all line integrals of $f(x,y)$ over all lines $L$:

	\begin{equation}
		\check{f} = \int_Lf(x,y)ds,
	\end{equation}
	\label{def:radon}

For lines with a certain angle $\phi$ with respect to the original reference system, the rotated coordinate system is written as

\[
\left(
\begin{array}{c}
x\\
y
\end{array}
\right) =
\left(
\begin{array}{cc}
\cos\phi&-\sin\phi\\
\sin\phi&\cos\phi
\end{array}
\right)
\left(
\begin{array}{c}
p\\
s
\end{array}
\right),
\]

\noindent accordingly, $\check{f}$ can be written explicitly:

\begin{equation}
\label{eq:radon}
\check{f}(p,\phi) = \int_{-\infty}^{\infty}f(p\cos\phi - s\sin\phi, p\sin\phi + s\cos\phi)ds.
\end{equation}

When the function $\check{f}(p,\phi)$ is known for all the $(p,\phi)$, the Radon transform of the function $f(x,y)$ is known. The space of the coordinates $(p,\phi)$ is known as Radon space, and its sampling is known as Sinogram.

The Radon transform offers a good description of how Gaia rectangular pixels and windowing scheme transforms the real continuous two-dimensional data into the Gaia's one-dimensional observations. In fact, for extended objects with radii smaller than half the sky mapper windows (thus fully contained in the transmitted windows), Gaia's observations are actually samplings of the object's signal in the Radon space.

To construct of the forward model, first the theoretical 2D brightness profile of the object is sampled on a sub-pixel-sized array. Then, this sampled signal is convolved with Gaia's PSF. Afterwards, given the scan angles in which the object was observed by Gaia, or the object's position in the sky, the numeric Radon transform of the resulting array is computed. Finally, the binning of the sky mapper and astro field windows are applied.

The fitting is then performed as the solution for $\min_{\mathbf{p}} \left|\left| f (\mathbf{p}) - \mathbf{d} \right|\right|_2$, where $\mathbf{p}$ is the vector with the model parameters at the solution, $f (\mathbf{p})$ is the forward model computed with the parameters $\mathbf{p}$ and $\mathbf{d}$ are the window data.

This optimization problem is solved with a hybrid optimizer, in a two-step strategy. First an approximate global optimization is performed, to locate the region of the parameter-space where the L2-norm's global optimum is likely to be found. Then, a local optimization is applied to accelerate the convergence. Moreover, adopting a local optimizer allows one to estimate the optimization errors at the solution.

For the mission data processing pipeline implementation, we currently plan to adopt the cross-entropy method \citep{YRubinstein:2004p9187} for the global optimization and the well-known quasi-newton BFGS method, or Broyden-Fletcher-Goldfarb-Shanno \citep{Broyden:1970p7232, Fletcher:1970p9063, Goldfarb:1970p7230, Shanno:1970p7231}, for the local optimization. However, for this conceptual analysis we adopted the RGenoud optimizer \citep{Mebane:2011p9381}, which uses a genetic algorithm for the global step and BFGS for the local step.

To test the profile fitting, we used SM windows. These are low-resolution data (236 mas/sample), but they allow a global analysis of the object to be performed, since they are the largest windows with 20 samples/window for all objects with $G>13$, and thus are the Gaia observations that will cover the largest area around the detected sources. Moreover, because of the low-resolution of the SM windows, we allowed a variation of the profile intensities, radii, ellipticities, and the background during this test, but we did not allow variations of the S\'ersic index (assumed to be $n=4$ for a de Vaucouleurs law) and of the $C_0$ parameter, because they should require higher-resolution data (from AF windows).

\begin{figure}
   \centering
  \begin{center}
  \subfloat[]{
    \includegraphics[width=0.49\linewidth, trim=0cm 0.5cm 0cm 0.5cm, clip=true]{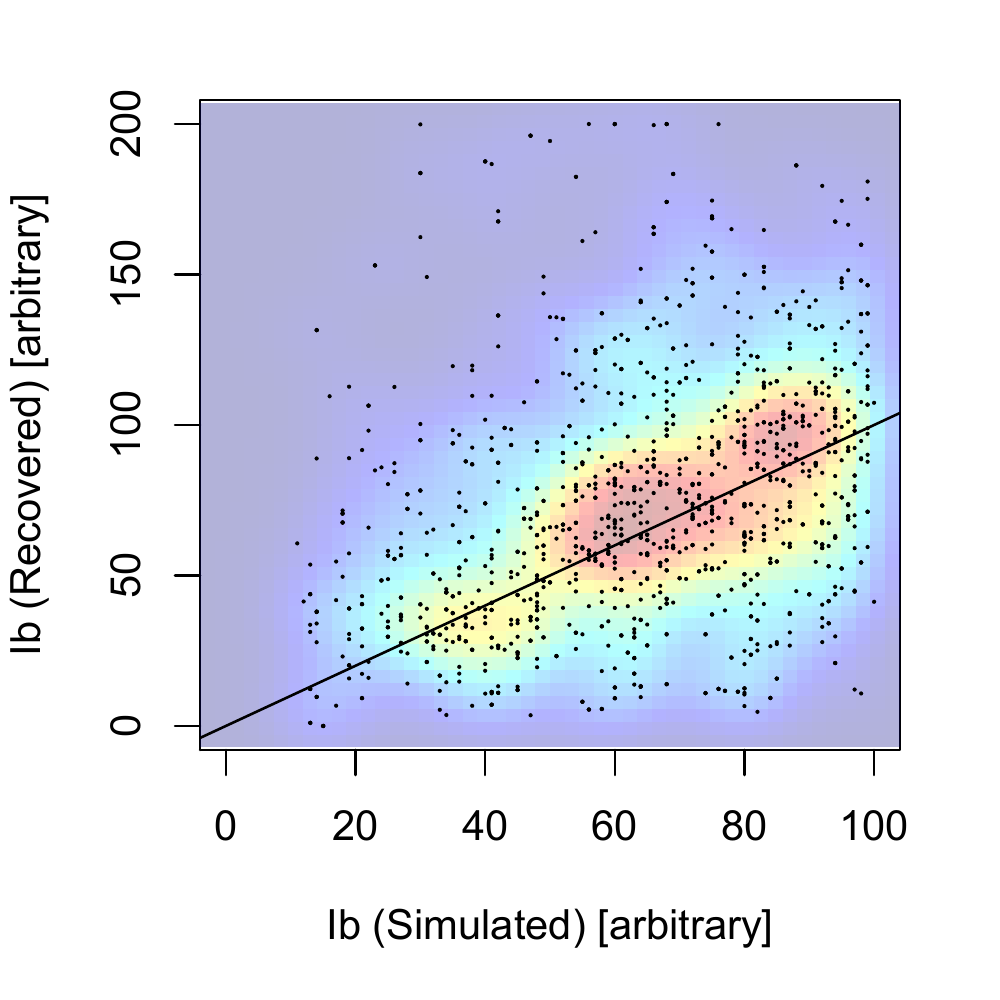}
  }
  \subfloat[]{
    \includegraphics[width=0.49 \linewidth, trim=0cm 0.5cm 0cm 0.5cm, clip=true]{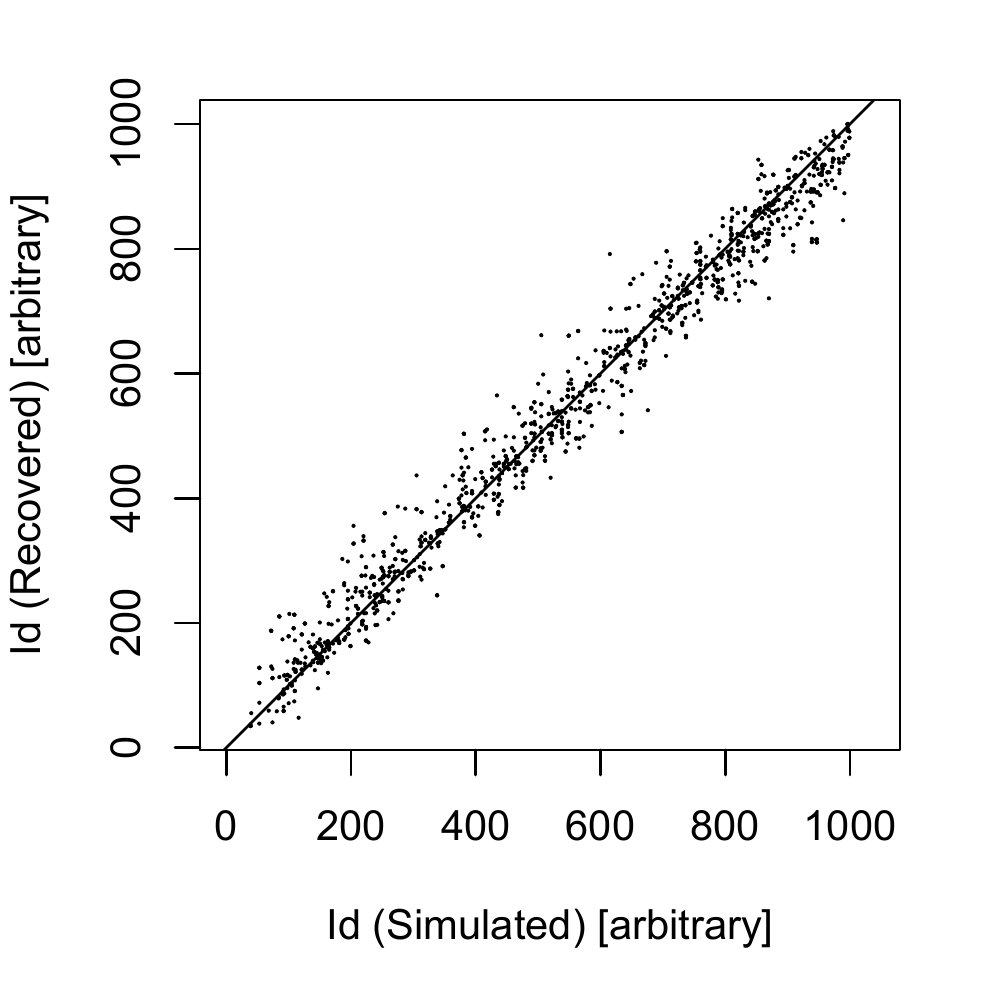}
  }\\
  \subfloat[]{
    \includegraphics[width=0.49 \linewidth, trim=0cm 0.5cm 0cm 0.5cm, clip=true]{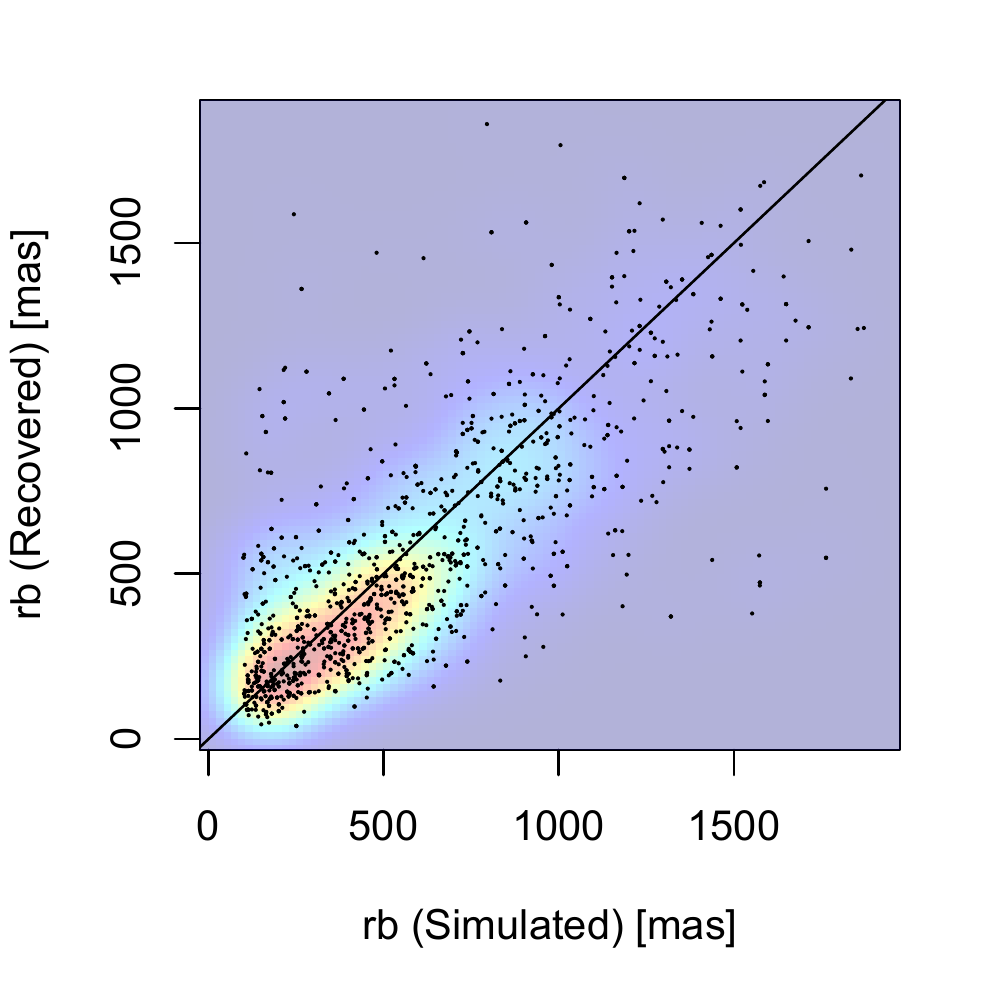}
  }
  \subfloat[]{
    \includegraphics[width=0.49 \linewidth, trim=0cm 0.5cm 0cm 0.5cm, clip=true]{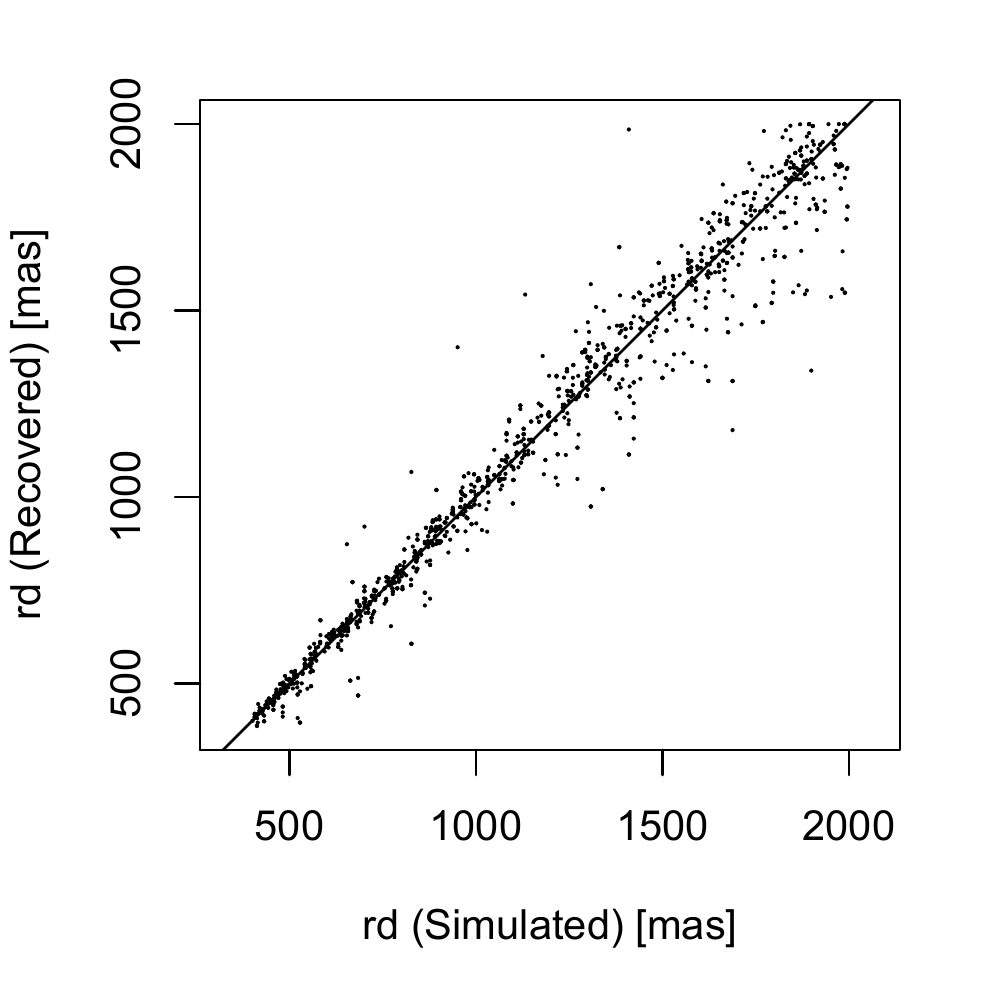}
  }\\
  \subfloat[]{
    \includegraphics[width=0.49 \linewidth, trim=0cm 0.5cm 0cm 0.5cm, clip=true]{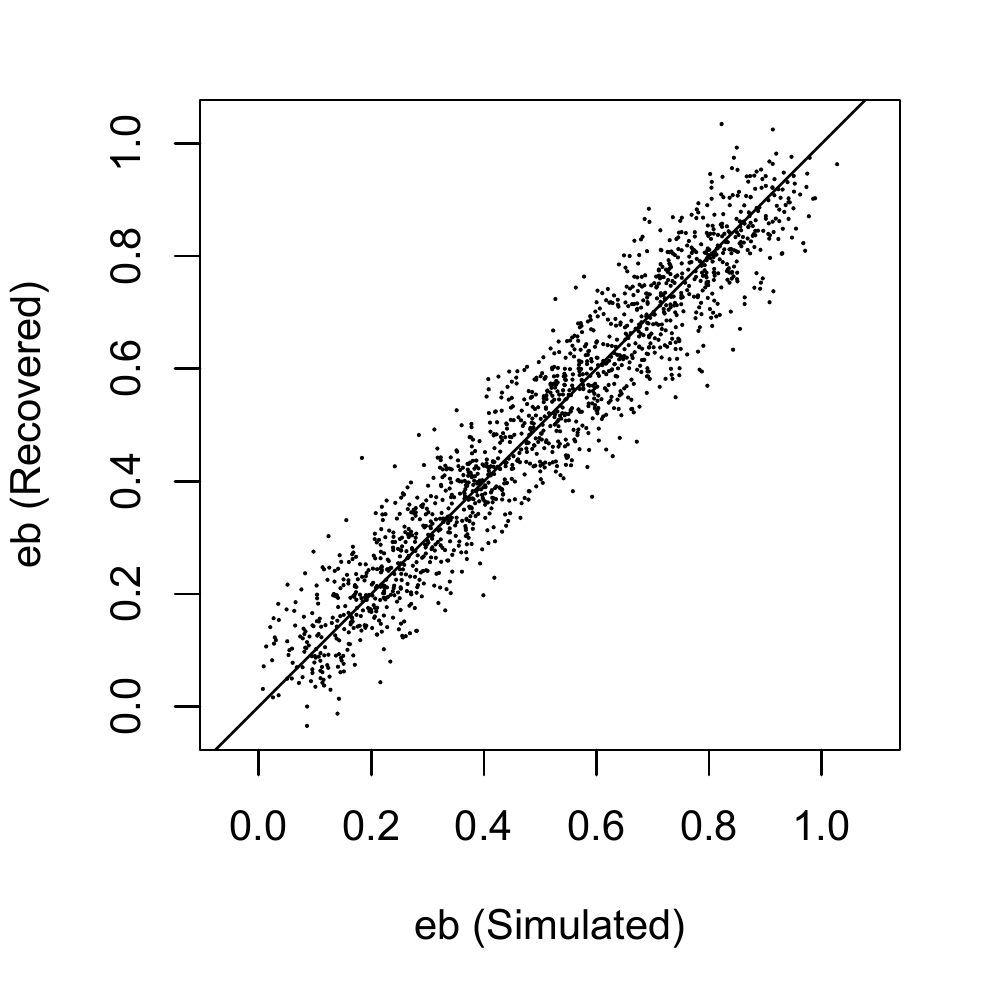}
  }
  \subfloat[]{
    \includegraphics[width=0.49 \linewidth, trim=0cm 0.5cm 0cm 0.5cm, clip=true]{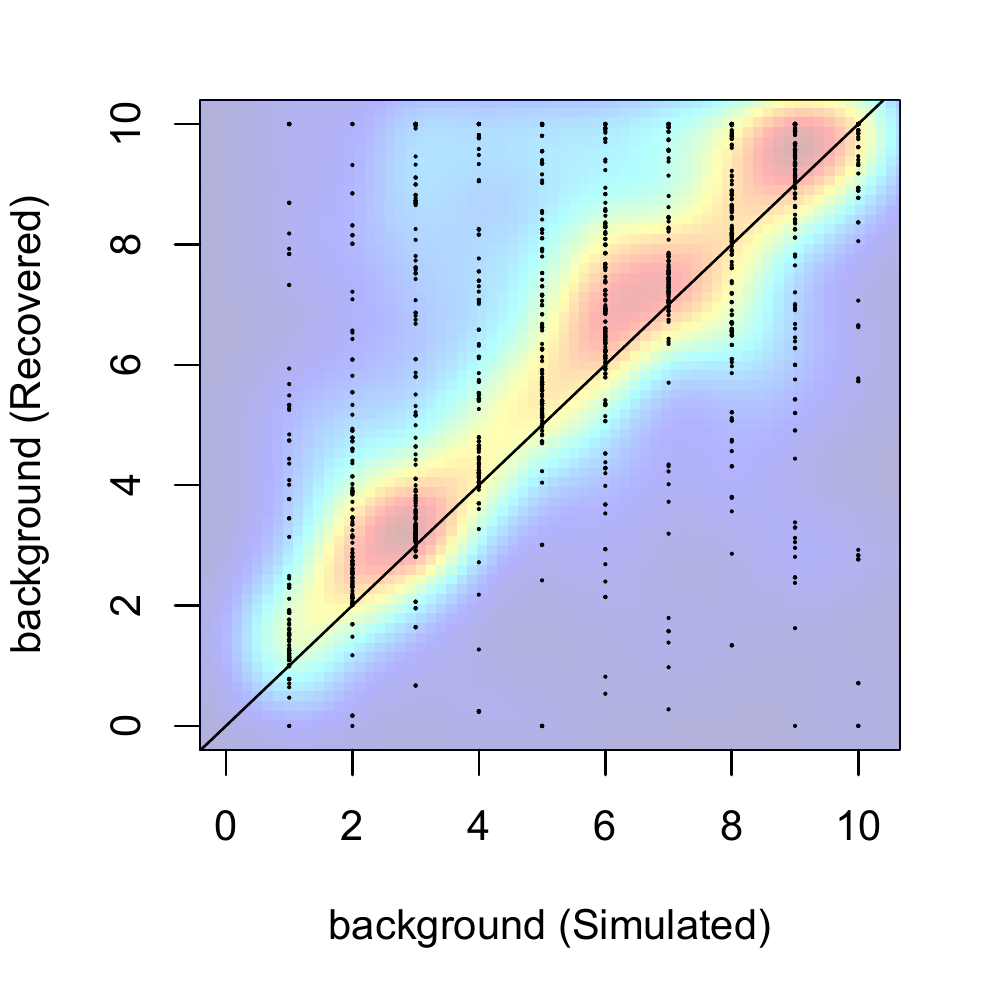}
 }
  \end{center}
      \caption{Simulated versus recovered profile parameters for 1600 randomly constructed galaxy profiles. The straight line indicates the ideal recovery. The background-color maps in the more disperse plots -- (a), (c) and (f) -- represent a kernel density estimation. In (e) a small Gaussian random offset with $\sigma=0.05$ was added to each point for readability.}
         \label{fig:ProfileFittingResults}
   \end{figure}

\begin{figure}
   \centering
  \begin{center}
  \subfloat[]{
    \includegraphics[width=0.49\linewidth, trim=0cm 0.5cm 0cm 0.5cm, clip=true]{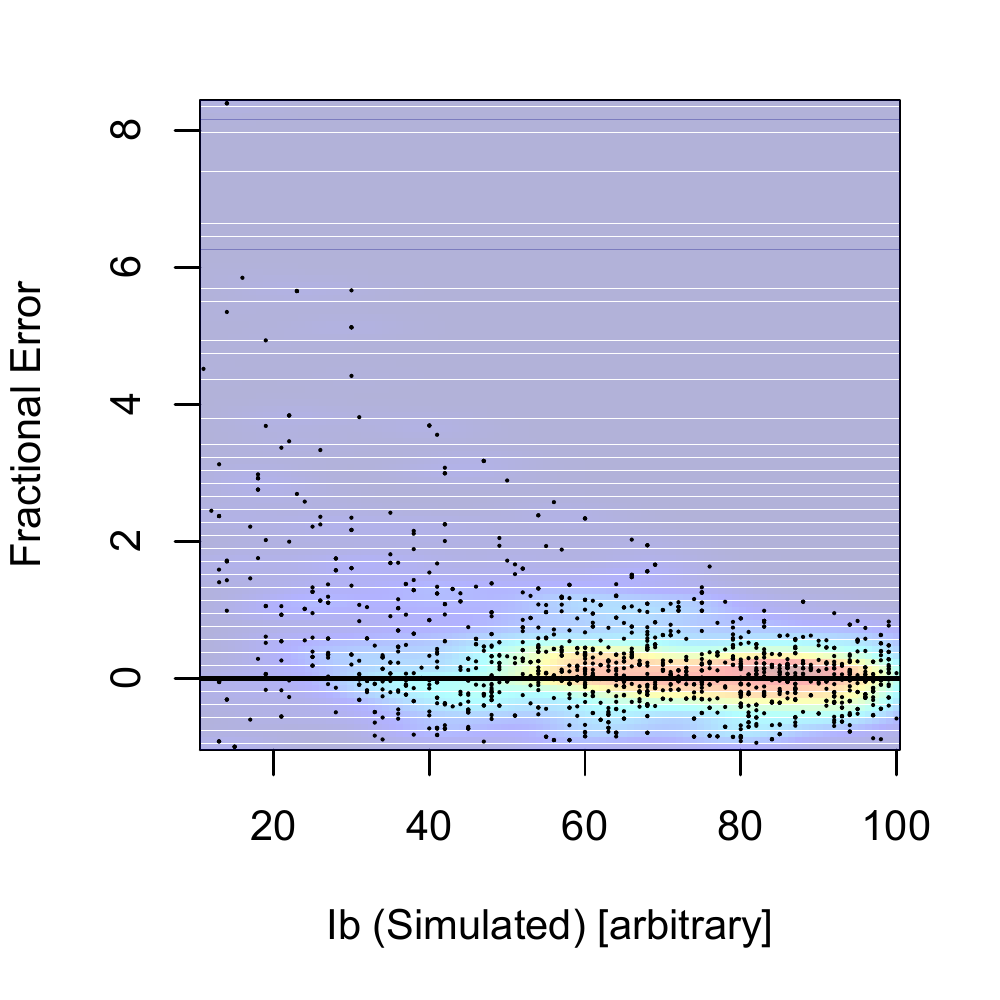}
  }
  \subfloat[]{
    \includegraphics[width=0.49 \linewidth, trim=0cm 0.5cm 0cm 0.5cm, clip=true]{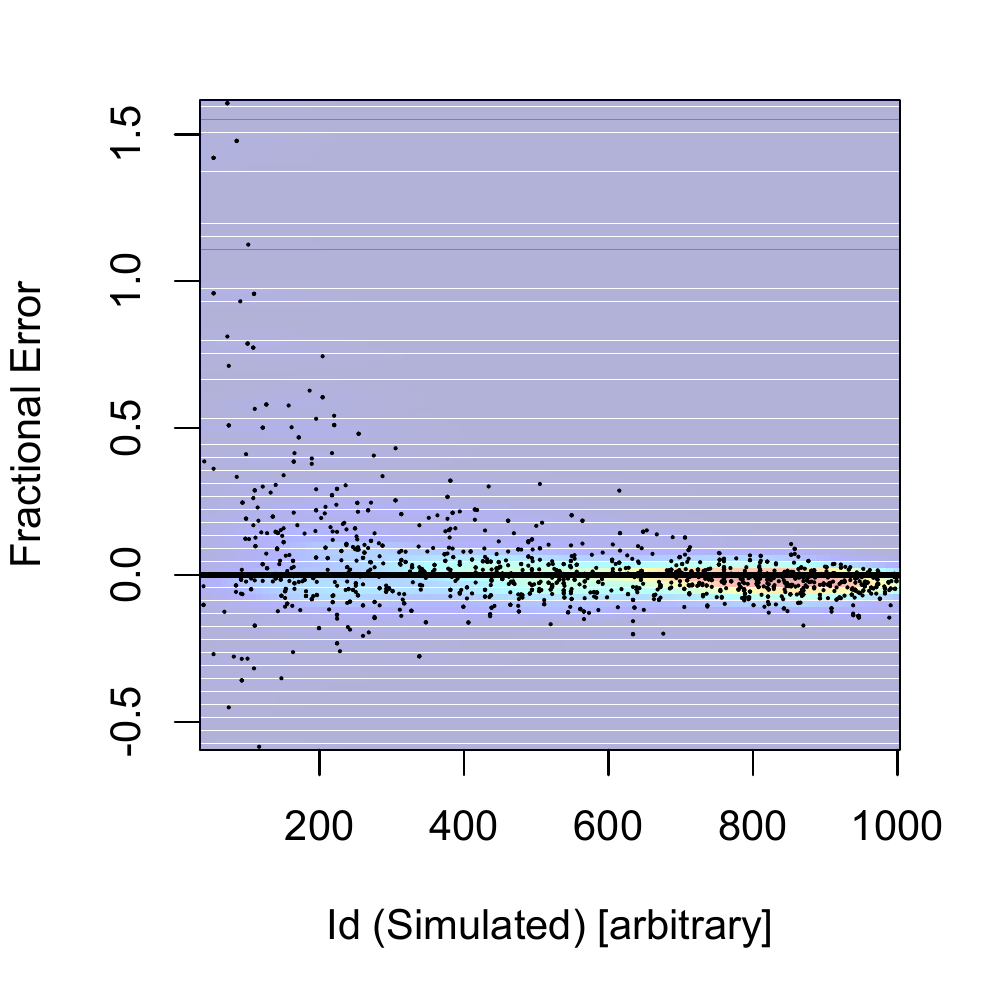}
  }\\
  \subfloat[]{
    \includegraphics[width=0.49 \linewidth, trim=0cm 0.5cm 0cm 0.5cm, clip=true]{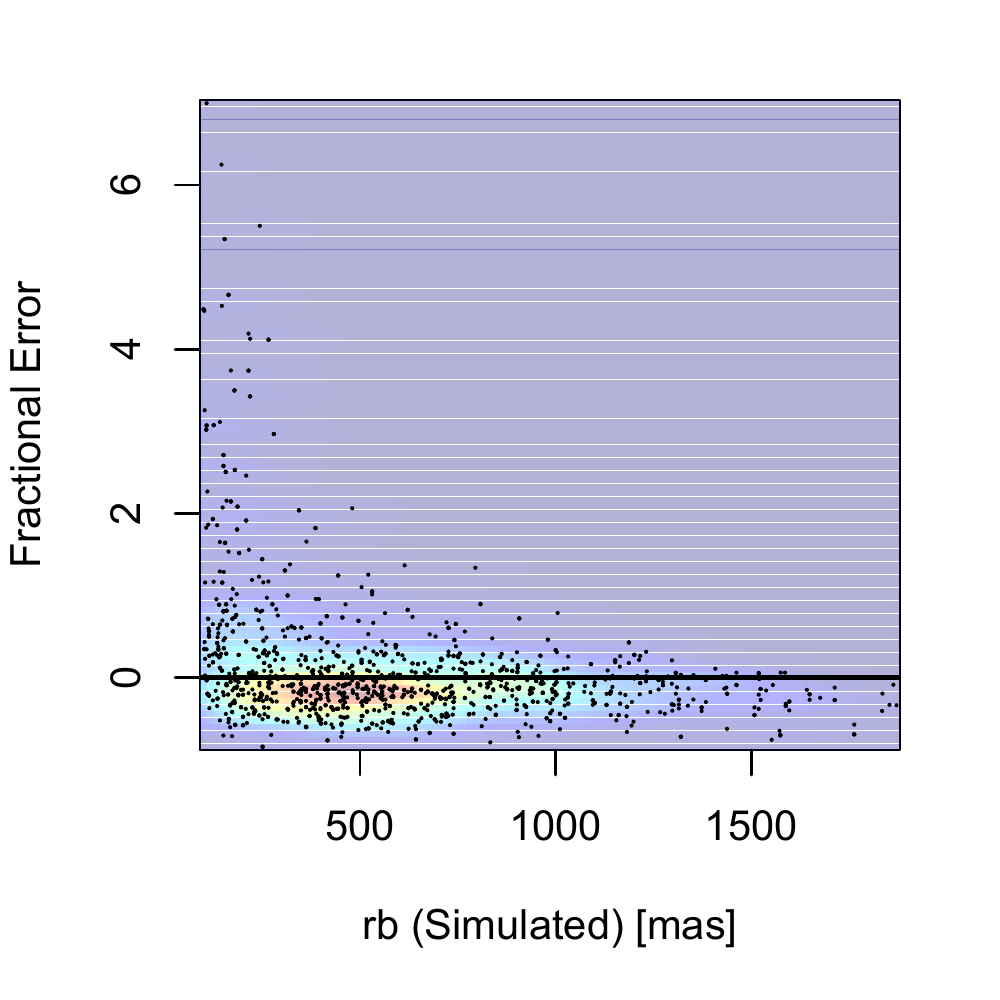}
  }
  \subfloat[]{
    \includegraphics[width=0.49 \linewidth, trim=0cm 0.5cm 0cm 0.5cm, clip=true]{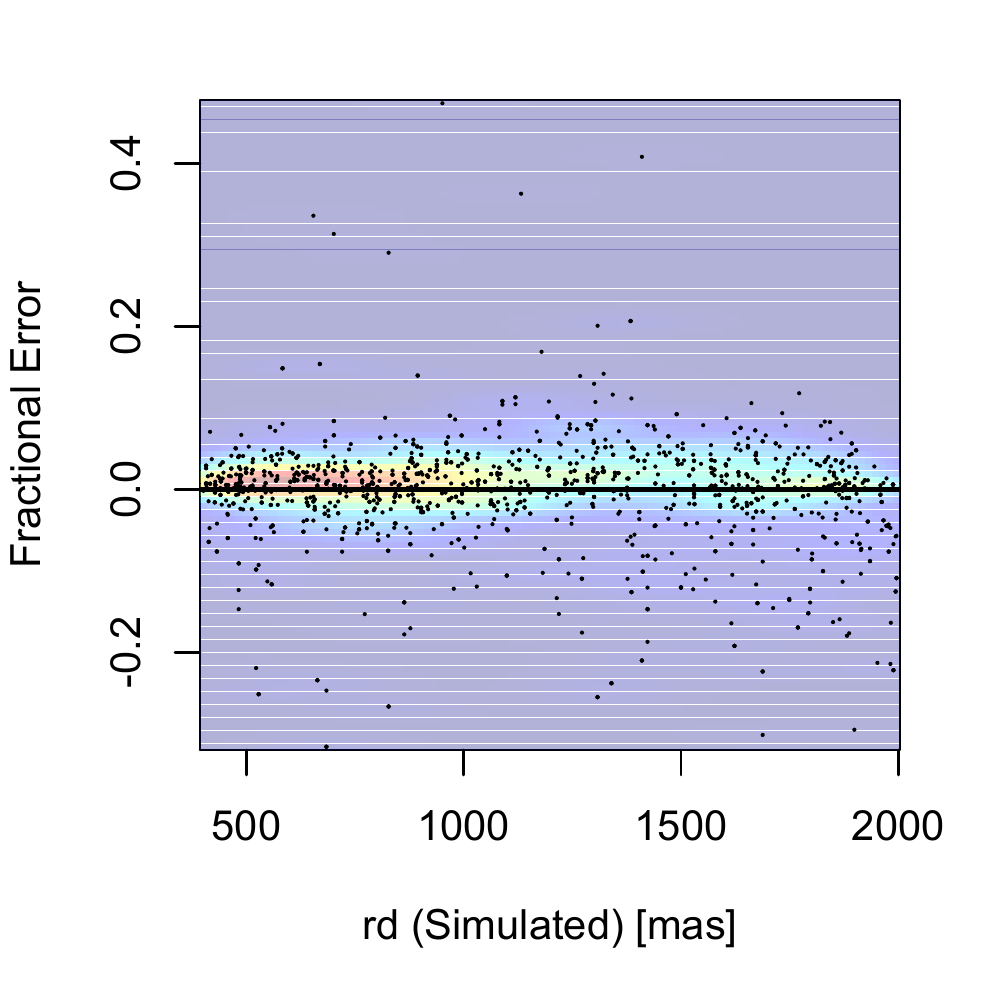}
  }\\
  \subfloat[]{
    \includegraphics[width=0.49 \linewidth, trim=0cm 0.5cm 0cm 0.5cm, clip=true]{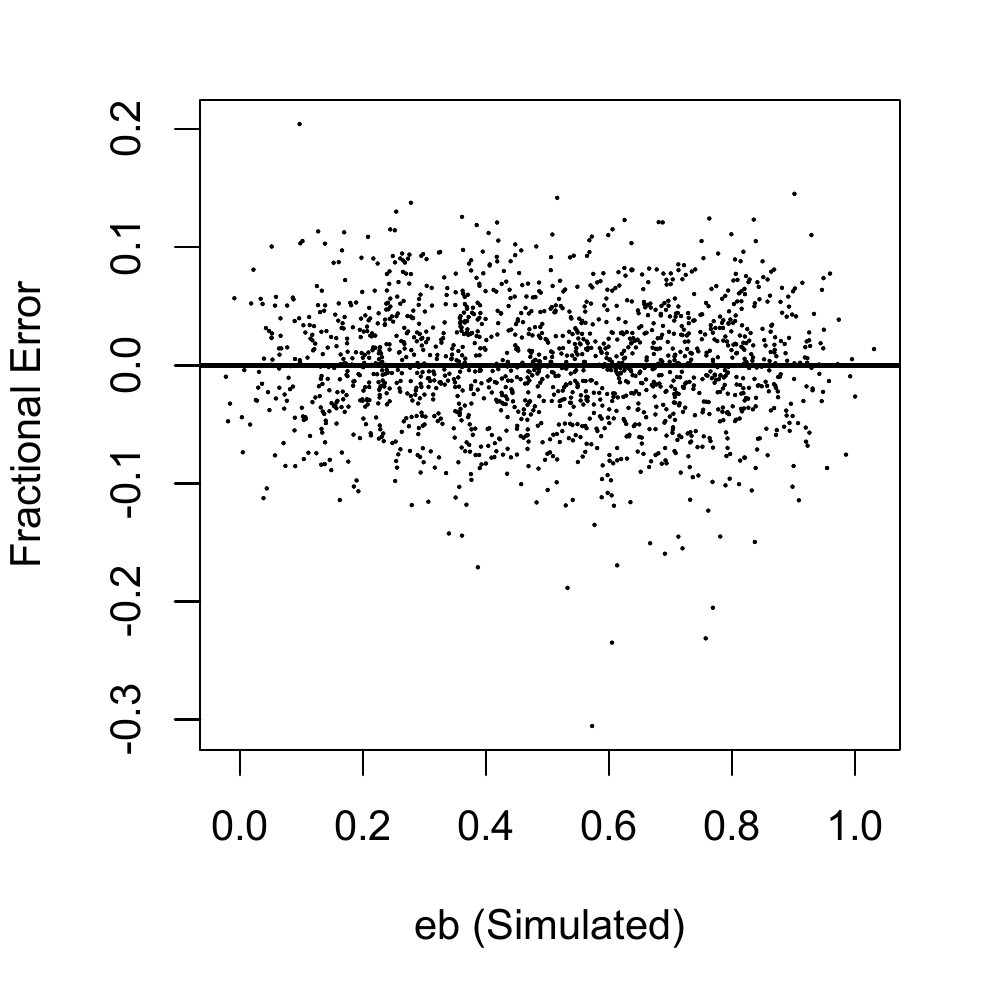}
  }
  \subfloat[]{
    \includegraphics[width=0.49 \linewidth, trim=0cm 0.5cm 0cm 0.5cm, clip=true]{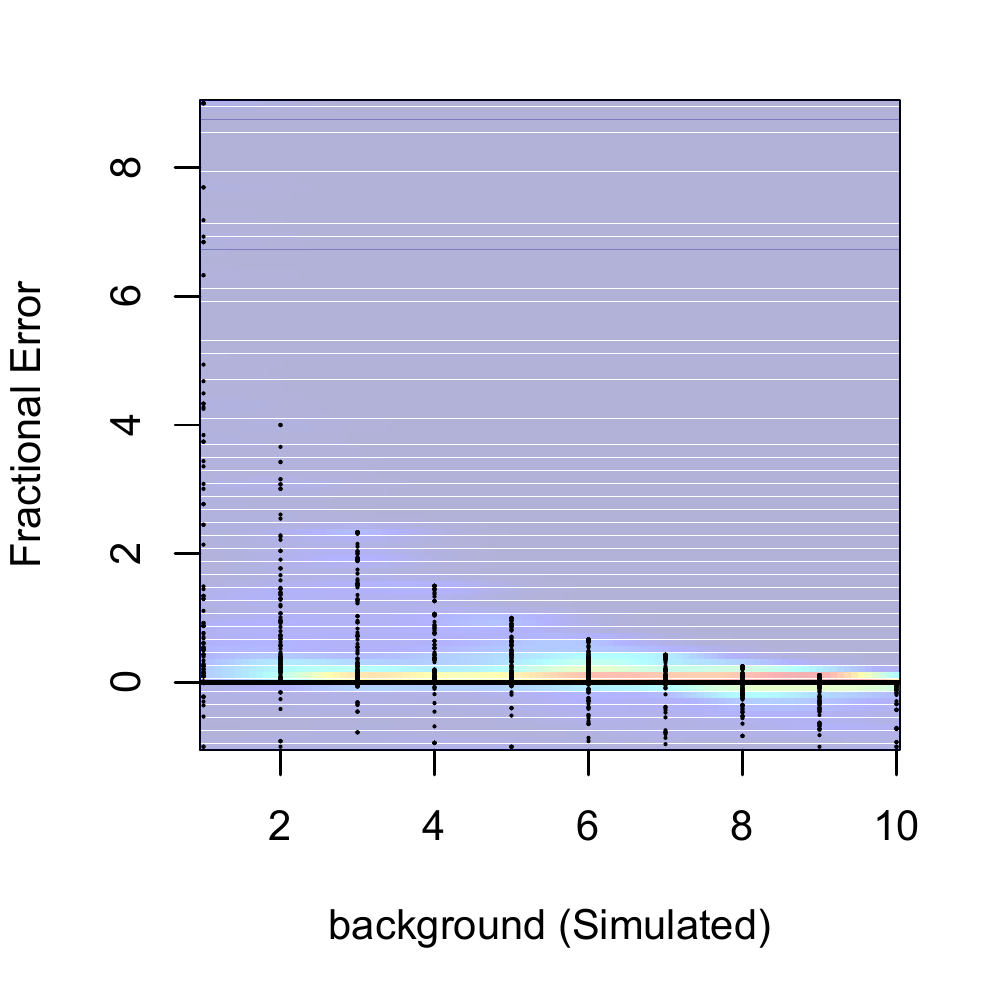}
 }
  \end{center}
      \caption{As Fig. 6, but representing fractional errors with the form $(\mathrm{Par}_{\mathrm{rec}}-\mathrm{Par}_{\mathrm{sim}})/\mathrm{Par}_{\mathrm{sim}}$ for the recovered parameters after the profile fitting using the Radon transform.}
         \label{fig:ProfileFittingResultsErrors}
   \end{figure}

The test comprised the simulation of 1600 galaxies, which were created from the most general profile, composed by bulge and disk components. As this test assesses the recovery of simulated profiles parameters, we simulated these galaxies directly using their analytical light profiles. The alternative solution would be the adoption of template images transformed by MAGIL, but this would restrict the coverage of the parameter space because the underlying image library has a finite size. The profile parameters of both components were randomly chosen within following intervals: $r_d \in [ 400; 2000]$ mas, $r_b \in [100; r_d]$ mas, $I_b \in [5\times \mathrm{background}; 100]$ arbitrary units, $I_d \in [I_b;1000]$ arbitrary units, $e \in [0.1;0.9]$ and $\mathrm{background} \in [1; 10]$ arbitrary units. These boundaries were selected to produce representative characteristics of objects this method may possibly encounter during the data treatment. The results of the profile fitting procedure for all simulated galaxies are presented in Fig. \ref{fig:ProfileFittingResults}, and the fractional errors, $(\mathrm{Par}_{\mathrm{rec}}-\mathrm{Par}_{\mathrm{sim}})/\mathrm{Par}_{\mathrm{sim}}$, in Fig.  \ref{fig:ProfileFittingResultsErrors}.\footnote{We note that no cuts on the goodness-of-fit value were applied, and all results are represented.}

From these results it is possible to notice that statistically, the simulated parameters were successfully recovered by the fitting procedure. However, there is a broad spread in the values obtained for the bulge parameters, as can be seen in Figs. \ref{fig:ProfileFittingResults} (a) and (c) and their respective fractional errors, represented in Figs. \ref{fig:ProfileFittingResultsErrors}. The median and median absolute deviation values of the difference between the recovered and simulated parameters, representing estimates for the systematic errors and random errors, respectively, are $\mathrm{med}(Ib)=11\pm53\%$ and $\mathrm{med}(rb)=-9\pm36\%$.

The broad spread observed on these values might arise because during this analysis the fitting was performed taking into account SM data only. Meanwhile, the bulge signal will be mainly present in the inner parts of the object, which are not well sampled by the adopted windows. In the performed simulations most of the bulge signal is within two or three samples only (each sky mapper sample is $\sim$200 mas wide). Furthermore, to compensate the error in the fitted $r_b$, the system tends to systematically overestimate the value of the bulge intensity at the fitted radius $I_b$. To correct for this problem, the implementation of the final data reduction pipeline is taking into account the higher-resolution data from the AF windows although they are more localized.

On the other hand, the disk profile parameters were well recovered, with a much more narrow spread, even when considering SM only data -- this can be seen in Figs. \ref{fig:ProfileFittingResults} (b) and (d) and the respective errors in Figs. \ref{fig:ProfileFittingResultsErrors}. The disk profile is well sampled by the SM data, since it spreads through a larger region due to its slower exponential gradient. The recovered disk parameters are quite linear, with a relatively low dispersion and without significant systematic trends: $\mathrm{med}(Id)=-1\pm7\%$ and $\mathrm{med}(rd)=1\pm4\%$.

The results also show that the galaxy ellipticities are recovered with a very low dispersion: since we allowed for a 0.1 resolution in fitting this parameters, the ellipticity is recovered at a level of accuracy better than $\sim 10\%$.  Because most of the points were superposed in Figs.~\ref{fig:ProfileFittingResults} (e) and~\ref{fig:ProfileFittingResultsErrors} (e), a Gaussian random offset with $\sigma = 0.05$ was added to each point to improve the visualization. Meanwhile, the background is slightly overestimated, $\mathrm{med}(background)=10\pm24\%$. Since for most simulated galaxies the background signal was much lower than the galaxy signal, the fitting procedure may be considering some of the flux from the bulge and disk components as background. However, during the nominal data analysis, the individual window backgrounds will be determined by a dedicated part of the Gaia data processing chain, and thus the background parameter will only be fitted at this stage if necessary as a mitigation measure, and should not interfere in the fitting in most cases.

The main conclusion obtained from this test is that using the proposed profile fitting procedure, based on a forward modeling approach, morphological parameters of galaxies may be recovered from one-dimensional Gaia-like data -- and even from its lowest spatial resolution observations.

\section{Conclusions}
When we started this work, no morphological analysis of the extended objects that will be observed by Gaia was planned in the mission. Thanks to the concepts described here, and to the implemented methods, we have shown that it is possible to use Gaia data to study the morphology of galaxies even when adopting only the satellite's lowest-resolution data.

We have presented the methods and test results for a conceptual pipeline that is at this moment being implemented in the Gaia Data Processing and Analysis Consortium to perform a morphological analysis of the galaxies observed by Gaia. This pipeline will be based on image reconstruction and on the measurement of image parameters to give hints about the object contained therein, in a support-vector-machine-based classifier and on a galaxy-light-profile-fitting (and thus bulge/disk decomposition) through forward modeling, using the Radon space and a hybrid optimizer applied to 1 D Gaia observations.

To validate the classification step, we adopted a subset of galaxies from the Hubble Deep Field North, that were previously visually classified. Using human classification as a benchmark, the success rate obtained by the system for early-types/spiral-irregular types is $\sim85\%$, while for the three morphological types it is $\sim81\%$, $\sim64\%$ and $\sim62\%$.

When the conceptual pipeline was applied to images reconstructed with the ShuffleStack  and FastStack methods from GIBIS simulated data, the results showed that there is still some room for improvement in the reconstruction methods. However, even adopting the less favorable method currently available, ShuffleStack, a fair  success rate for classifying two classes (ellipticals/spirals-irregulars) types is obtained at $\sim83\%$, as well as for three morphological classes (ellipticals, spirals, irregulars/peculiars), which present success rates of $79\%$, $56\%$, and $74\%$.

The final, and arguably the most important, step is the profile fitting. Results obtained from simulated observations of Gaia low-resolution data indicate that the morphological parameters of the light profile are recovered with errors (systematic$\pm$random) at the following levels: -9$\pm$36\% for the bulge radius, 11$\pm$53\% for the bulge intensity, 1$\pm$4\% for the disk radius, and -1$\pm$7\% for the disk intensity.

The design of the pipeline that will be actually used during the mission data processing, its engineering aspects and the performance results for a global analysis from low- and high-resolution data (SM and AF windows), which will allow for a better assessment of the bulge, will be presented in a future work.

Finally, based on the results presented here, which were derived exclusively from low-resolution (highly binned) SM data, we conclude that it will be possible to harness the information of many Gaia observations for the sake of extragalactic science. 

Gaia was not designed for any kind of morphological analysis of extended objects. However, saving these data will allow studying the morphology of about one million galaxies whose structures can only be probed from space or by adopting adaptive optics, thus somewhat expanding the horizons of the already comprehensive ESA Gaia space mission.

 \begin{acknowledgements}
This work was supported by the Brazilian agencies FAPESP, the Brazilian-French cooperation agreement CAPES-COFECUB, the French agencies CNRS, CNES, the {\it Action Sp\'ecifique Gaia}, and the Portuguese agency FCT (SFRH/BPD/74697/2010, PDCTE/CTE-AST/81711/2003, PTDC/CTE-SPA/118692/2010).

We also thank the French CNES and the Brazilian INCT-A for providing computational resources, A. Bijaoui, C. Dollet, and C. Pereyga for the pioneer work on the image reconstruction for Gaia and for useful discussions, D. Harrisson for the adopted reconstruction algorithms, C. Babusiaux, X. Luri, and the Gaia Data Processing and Analysis Consortium Coordination Unit 2 - Simulations for the {\it Gaia Instrument and Basic Image Simulator} and finally, the CNES CU4 Team (B. Frezouls, G. Prat, K.-C. Pham, R. Pedrosa, V. Valette) for all the support relative to database and code development. Finally, we thank the referee for helpful suggestions.

\end{acknowledgements}

\bibliographystyle{aa}
\bibliography{AA_2012_19697.bib}

\begin{thebibliography}{56}
\expandafter\ifx\csname natexlab\endcsname\relax\def\natexlab#1{#1}\fi
\expandafter\ifx\csname url\endcsname\relax
  \def\url#1{{\tt #1}}\fi
\expandafter\ifx\csname urlprefix\endcsname\relax\def\urlprefix{URL }\fi

\bibitem[{Abraham et~al.(1994)Abraham, Valdes, Yee, \& van~den
  Bergh}]{Abraham:1994p4126}
Abraham R.G., Valdes F., Yee H.K.C., van~den Bergh S., 1994, \apj, 432, 75

\bibitem[{Abraham et~al.(1996)Abraham, van~den Bergh, Glazebrook
  et~al.}]{Abraham:1996p4217}
Abraham R.G., van~den Bergh S., Glazebrook K., et~al., 1996, \apjs, 107, 1

\bibitem[{Abraham et~al.(2003)Abraham, van~den Bergh, \&
  Nair}]{Abraham:2003p4894}
Abraham R.G., van~den Bergh S., Nair P., 2003, \apj, 588, 218

\bibitem[{{Ahn} et~al.(2012){Ahn}, {Alexandroff}, {Allende Prieto}
  et~al.}]{ahn2012}
{Ahn} C.P., {Alexandroff} R., {Allende Prieto} C., et~al., 2012, \apjs, 203, 21

\bibitem[{Aime et~al.(1978)Aime, Ricort, \& Harvey}]{Aime1978}
Aime C., Ricort G., Harvey J., 1978, \apj, 221, 362

\bibitem[{Athanassoula et~al.(1990)Athanassoula, Morin, Wozniak
  et~al.}]{Athanassoula:1990p6909}
Athanassoula E., Morin S., Wozniak H., et~al., 1990, \mnras, 245, 130

\bibitem[{Babusiaux et~al.(2009)Babusiaux, Sartoretti, Leclerc, Ch{\'e}reau, \&
  Weiler}]{GIBIS7}
Babusiaux C., Sartoretti P., Leclerc N., Ch{\'e}reau F., Weiler M., 2009,
  GAIA-C2-SP-OPM-CB-003

\bibitem[{{Beckwith} et~al.(2006){Beckwith}, {Stiavelli}, {Koekemoer}
  et~al.}]{Beckwith2006}
{Beckwith} S.V.W., {Stiavelli} M., {Koekemoer} A.M., et~al., 2006, \aj, 132,
  1729

\bibitem[{{Blanton} \& {Moustakas}(2009)}]{Blanton2009}
{Blanton} M.R., {Moustakas} J., 2009, \araa, 47, 159

\bibitem[{Blanton et~al.(2001)Blanton, Dalcanton, Eisenstein
  et~al.}]{Blanton:2001p6552}
Blanton M.R., Dalcanton J., Eisenstein D., et~al., 2001, \aj, 121, 2358

\bibitem[{Broyden(1970)}]{Broyden:1970p7232}
Broyden C., 1970, IMA Journal of Applied Mathematics, 6, 76

\bibitem[{Capaccioli(1989)}]{Capaccioli:1989p7121}
Capaccioli M., 1989, Proc. of the Conference ``The world of galaxies'', 208

\bibitem[{Case et~al.(2009)Case, Cherry, Ling, Shimizu, \& Wheaton}]{Case2009}
Case G.L., Cherry M.L., Ling J.C., Shimizu T., Wheaton W.A., 2009,
  arXiv:0912.3815v1

\bibitem[{Chang \& Lin(2011)}]{CC01a}
Chang C.C., Lin C.J., 2011, ACM Transactions on Intelligent Systems and
  Technology, 2, 27:1

\bibitem[{Conselice(2003)}]{Conselice:2003p5139}
Conselice C.J., 2003, \apjs, 147, 1

\bibitem[{Conselice(2006)}]{Conselice:2006p3903}
Conselice C.J., 2006, \mnras, 373, 1389

\bibitem[{Cortes \& Vapnik(1995)}]{Cortes:1995p9085}
Cortes C., Vapnik V., 1995, Machine learning, 20, 273

\bibitem[{Deans(1983)}]{Deans1983}
Deans S., 1983, The Radon transform and some of its applications,
  Wiley-Interscience, New York

\bibitem[{Doi et~al.(1993)Doi, Fukugita, \& Okamura}]{Doi:1993p5085}
Doi M., Fukugita M., Okamura S., 1993, \mnras, 264, 832

\bibitem[{Dollet et~al.(2005)Dollet, Bijaoui, \& Mignard}]{Dollet2005}
Dollet C., Bijaoui A., Mignard F., 2005, Proc. of the Gaia Symposium ``The
  Three-Dimensional Universe with Gaia'' (ESA SP-576), 576, 429

\bibitem[{Fletcher(1970)}]{Fletcher:1970p9063}
Fletcher R., 1970, The Computer Journal, 13, 317

\bibitem[{Frei et~al.(1996)Frei, Guhathakurta, Gunn, \& Tyson}]{Frei:1996p5214}
Frei Z., Guhathakurta P., Gunn J.E., Tyson J.A., 1996, \aj, 111, 174

\bibitem[{Gavras et~al.(2010)Gavras, Krone-Martins, Ducourant, \&
  Sinachopoulos}]{Gavras_MAGIL_2010}
Gavras P., Krone-Martins A., Ducourant C., Sinachopoulos D., 2010,
  GAIA-C4-SP-NOA-PG-001-1

\bibitem[{{Giavalisco} et~al.(2004){Giavalisco}, {Ferguson}, {Koekemoer}
  et~al.}]{Giavalisco2004}
{Giavalisco} M., {Ferguson} H.C., {Koekemoer} A.M., et~al., 2004, \apjl, 600,
  L93

\bibitem[{Goldfarb(1970)}]{Goldfarb:1970p7230}
Goldfarb D., 1970, Mathematics of Computation, 24, 23

\bibitem[{Graham \& Driver(2005)}]{Graham:2005p7165}
Graham A.W., Driver S.P., 2005, \pasa, 22, 118

\bibitem[{Harrison(2011)}]{Harrison:2011p8791}
Harrison D.L., 2011, Experimental Astronomy, 31, 157

\bibitem[{Huertas-Company et~al.(2008)Huertas-Company, Rouan, Tasca, Soucail,
  \& F{\`e}vre}]{HuertasCompany:2008p5093}
Huertas-Company M., Rouan D., Tasca L., Soucail G., F{\`e}vre O.L., 2008, \aap,
  478, 971

\bibitem[{Jordi et~al.(2010)Jordi, Gebran, Carrasco et~al.}]{Jordi:2010p7353}
Jordi C., Gebran M., Carrasco J.M., et~al., 2010, \aap, 523, 48

\bibitem[{Krone-Martins et~al.(2008)Krone-Martins, Ducourant, \&
  Teixeira}]{KroneMartins:2008p2436}
Krone-Martins A., Ducourant C., Teixeira R., 2008, Proceedings of the
  International Conference: ``Classification and Discovery in Large
  Astronomical Surveys''. AIP Conference Proceedings., 1082, 151

\bibitem[{Krone-Martins et~al.(2012)Krone-Martins, Ducourant, \&
  Teixeira}]{AKM-004}
Krone-Martins A., Ducourant C., Teixeira R., 2012, GAIA-C4-TN-FCUL-AKM-004

\bibitem[{{Leauthaud} et~al.(2007){Leauthaud}, {Massey}, {Kneib}
  et~al.}]{Leauthaud2007}
{Leauthaud} A., {Massey} R., {Kneib} J.P., et~al., 2007, \apjs, 172, 219

\bibitem[{Lin et~al.(2008)Lin, Lee, Chen, \& Tseng}]{Lin:2008p1505}
Lin S.W., Lee Z.J., Chen S.C., Tseng T.Y., 2008, Applied Soft Computing, 8,
  1505

\bibitem[{Lindegren(2010)}]{Lindegren:2010p9251}
Lindegren L., 2010, Relativity in Fundamental Astronomy: Dynamics, 261, 296

\bibitem[{Lotz et~al.(2004)Lotz, Primack, \& Madau}]{Lotz:2004p5429}
Lotz J.M., Primack J., Madau P., 2004, \aj, 128, 163

\bibitem[{Mebane \& Sekhon(2011)}]{Mebane:2011p9381}
Mebane W., Sekhon J., 2011, Journal of Statistical Software, 42, 1

\bibitem[{Mignard et~al.(2008)Mignard, Bailer-Jones, Bastian
  et~al.}]{Mignard:2008p9245}
Mignard F., Bailer-Jones C., Bastian U., et~al., 2008, A Giant Step: from
  Milli- to Micro-arcsecond Astrometry, 248, 224

\bibitem[{Mo et~al.(2011)Mo, van~den Bosh, \& White}]{Mo2011}
Mo H., van~den Bosh F., White S., 2011, Galaxy Formation and Evolution,
  Cambridge Univ. Press., New York

\bibitem[{Nurmi(2005)}]{Nurmi:2005p9057}
Nurmi P., 2005, The Three-Dimensional Universe with Gaia, 461

\bibitem[{Parks(1986)}]{Parks1986}
Parks P.C., 1986, IEEE Transactions on Medical Imaging, MI-5, 170

\bibitem[{Perryman et~al.(2001)Perryman, de~Boer, Gilmore
  et~al.}]{Perryman:2001p3838}
Perryman M.A.C., de~Boer K.S., Gilmore G., et~al., 2001, \aap, 369, 339

\bibitem[{Peyrega(2007)}]{Pereyga2007}
Peyrega C., 2007, Reconstruction d'images des champs de la mission GAIA de
  l'ESA {\`a} partir des fen{\^e}tres transmises sur Terre, MSc dissertation
  (adv. A. Bijoaui), Observatoire de la C{\^o}te d'Azur

\bibitem[{Radon(1917)}]{Radon1917}
Radon J., 1917, Berichte der S{\"a}chsischen Akadamie der Wissenschaft, 69, 262

\bibitem[{Robin et~al.(2012)Robin, Luri, Reyl{\'e} et~al.}]{Robin:2012p9531}
Robin A.C., Luri X., Reyl{\'e} C., et~al., 2012, \aap, 543, 100

\bibitem[{Rubinstein \& Kroese(2004)}]{YRubinstein:2004p9187}
Rubinstein R.Y., Kroese D.P., 2004, The cross-entropy method: a unified
  approach to combinatorial optimization, Springer Verlag

\bibitem[{{Scoville} et~al.(2007){Scoville}, {Abraham}, {Aussel}
  et~al.}]{scoville2007}
{Scoville} N., {Abraham} R.G., {Aussel} H., et~al., 2007, \apjs, 172, 38

\bibitem[{Shanno(1970)}]{Shanno:1970p7231}
Shanno D., 1970, Mathematics of Computation, 24, 647

\bibitem[{de~Souza et~al.(2013, in preparation)de~Souza, dos Anjos, Teixeira,
  Ducourant, \& Krone-Martins}]{deSouza2013}
de~Souza R.E., dos Anjos S., Teixeira R., Ducourant C., Krone-Martins A., 2013,
  in preparation

\bibitem[{Starck et~al.(2003)Starck, Donoho, \& Cand{\`e}s}]{Starck:2003p3898}
Starck J.L., Donoho D.L., Cand{\`e}s E.J., 2003, \aap, 398, 785

\bibitem[{{Tasca} et~al.(2009){Tasca}, {Kneib}, {Iovino} et~al.}]{Tasca2009}
{Tasca} L.A.M., {Kneib} J.P., {Iovino} A., et~al., 2009, \aap, 503, 379

\bibitem[{Touma(1997)}]{Touma1997}
Touma H., 1997, \apss, 258, 47

\bibitem[{{Tsalmantza} et~al.(2007){Tsalmantza}, {Kontizas}, {Bailer-Jones}
  et~al.}]{Tsalmantza2007}
{Tsalmantza} P., {Kontizas} M., {Bailer-Jones} C.A.L., et~al., 2007, \aap, 470,
  761

\bibitem[{{Tsalmantza} et~al.(2009){Tsalmantza}, {Kontizas}, {Rocca-Volmerange}
  et~al.}]{Tsalmantza2009}
{Tsalmantza} P., {Kontizas} M., {Rocca-Volmerange} B., et~al., 2009, \aap, 504,
  1071

\bibitem[{Vapnik(2000)}]{vapnik2000nature}
Vapnik V., 2000, {The Nature of Statistical Learning Theory}, Springer Verlag

\bibitem[{Williams et~al.(1996)Williams, Blacker, Dickinson
  et~al.}]{Williams1996}
Williams R.E., Blacker B., Dickinson M., et~al., 1996, \aj, 112, 1335

\bibitem[{Zhang et~al.(1993)Zhang, Fishman, Harmon, \& Paciesas}]{Zhang1993}
Zhang S.N., Fishman G.J., Harmon B.A., Paciesas W.S., 1993, \nat, 366, 245

\end{thebibliography}

\end{document}